\documentclass[aap,MSNbibl,seceqn,nameyear,dvips]{arximspdf}
\usepackage{mathbh}
\usepackage{graphicx}

\doi{10.1214/13-AAP958} 
\volume{24}
\issue{4}
\pubyear{2014}
\firstpage{1621}
\lastpage{1651}

\makeatletter
\newcommand{\widebar}{\overline}

\newcommand{\rright}{\right}
\newcommand{\lleft}{\left}
\newcommand{\rrvert}{\vert}
\newcommand{\llvert}{\vert}
\newtheorem{teo}{Theorem}[section]
\newtheorem{cor}[teo]{Corollary}
\newtheorem{prop}[teo]{Proposition}
\newtheorem{lemma}[teo]{Lemma}
\newproclaim{rem}{Remark}
\makeatother

\begin{document}
\begin{frontmatter}

\title{Extremal laws for the real Ginibre ensemble}
\runtitle{Real Ginibre ensemble}

\begin{aug}
\author[A]{\fnms{Brian} \snm{Rider}\corref{}\ead[label=e1]{Brian.Rider@temple.edu}\thanksref{t1}}
\and
\author[B]{\fnms{Christopher D.} \snm{Sinclair}\ead[label=e2]{csinclai@uoregon.edu}\thanksref{t2}}
\runauthor{B. Rider and C. D. Sinclair}
\affiliation{Temple University and University of Oregon}
\address[A]{Department of Mathematics\\
Temple University\\
Philadelphia, Pennsylvania 19122\\
USA\\
\printead{e1}} 
\address[B]{Department of Mathematics\\
University of Oregon\\
Eugene, Oregon 97403\\
USA\\
\printead{e2}}
\end{aug}
\thankstext{t1}{Supported in part by NSF Grant DMS-06-45756 as well as
Grant 229249 from the Simons Foundation.}
\thankstext{t2}{Supported in part by NSF Grant DMS-08-01243.}

\received{\smonth{10} \syear{2012}}
\revised{\smonth{7} \syear{2013}}

%
\begin{abstract}
The real Ginibre ensemble refers to the family of $n \times n$
matrices in which each entry is an independent Gaussian random
variable of mean zero and variance one. Our main result is that the
appropriately scaled spectral radius converges in law to a Gumbel
distribution as $n \rightarrow\infty$. This fact has been known to
hold in the complex and quaternion analogues of the ensemble for some
time, with simpler proofs. Along the way we establish a new form for
the limit law of the largest real eigenvalue.
\end{abstract}

%

\end{frontmatter}

\section{Introduction}\label{sec1}

Ginibre (\citeyear{MR0173726}) introduced the basic non-Hermitian
ensembles of random matrix theory. These are $n \times n$ matrices
$M$ comprised of independent (and standardized) real, complex or
quaternion Gaussian entries and are clear analogues of the Gaussian
orthogonal, unitary and symplectic ensembles $(\mathrm{G}\{\mathrm{O}/\mathrm{U}/\mathrm{S}\}\mathrm{E})$.

The results of \citet{MR0173726} include explicit formulas for the
joint density of eigenvalues of $M$, in both the complex and
quaternion cases. The real-entried case posed serious technical
hurdles, due largely to the fact that the real line itself receives
positive mass in this setting, and the determination of the joint
eigenvalue density remained open until \citet{MR1437734},
\citet{1991PhRvL67941L}. These
papers found conditional densities for the real Ginibre ensemble
eigenvalues, that is, formulas for the joint law given a predetermined
number of real eigenvalues. Even with these in hand, the expressions
proved sufficiently complicated that the finite-dimensional
correlation functions---the basic tool(s) required to obtain limit
theorems for local eigenvalue statistics---were only determined in
the last few years. \citet{borodin-2007} rigorously established that the
eigenvalues of the real Ginibre
ensemble form a Pfaffian point process: there are $2 \times2$ skew
matrix kernels for the real/real, complex/complex and real/complex
correlations from which the general $k$-point correlation is built as
a $2k \times2k$ Pfaffian. These formulas along with the connected
skew orthogonal polynomials discovered by
\citet{forrester-2007} allowed Borodin--Sinclair to derive the scaling
limits for the kernels at both the (real and complex) bulk and the
(real and complex) edge [\citet{borodin-2008}]. [We note that \citet{forrester-2007}
also presents the real/real and complex/complex correlations, as well
as some asymptotics
for the one and two-point functions.
Concurrently,
\citet{sommers-2007} reported the scaling limits of the kernels in the
bulk.]

Here we are after a scaling limit for the spectral radius in the real
Ginibre ensemble. Among other motivations, this may be viewed as one
possible refinement of the circular law. The latter refers to the fact
that the normalized counting measure of the scaled eigenvalues
converges (weakly almost surely) to the uniform measure on the unit
disk. This result has a rather long history, starting with the work of
\citet{MR773436} which was made rigorous by \citet{MR1428519},
and culminating in the universality (in terms of entry distributions)
theorems of \citet{MR2663633} and then \citet{MR2722794}. On a local
scale, considerable progress
has been made on the universality of the $n \uparrow\infty$ bulk
correlations (even all the way up to the edge) for both real and
complex entries; see
\citeauthor{2012arXiv12061449B} (\citeyear{2012arXiv12061449B,2012arXiv12063187B}), \citet{2012arXiv12061893T}.

Our main result is the following:

%
%
\begin{teo}
\label{mainthm}
Denote by $R_n$ the spectral radius of the real Ginibre ensemble, and
set $\gamma_n = \log(n/ (2\pi(\log n)^2) )$. Then, as $n \rightarrow
\infty$,
\[
\sqrt{ 4 \gamma_n } \biggl( R_n - \sqrt{n} - \sqrt{
\frac{\gamma_n}{4}} \biggr) \Rightarrow G,
\]
where $G$ is the Gumbel law with distribution function $F_G(t)= e^{-(1/2) e^{-t}}$.
\end{teo}

One can certainly adjust the scaling so that the limiting
distribution function takes the more standard form of $e^{-e^{-t}}$.
It is written this way for comparison: at the same scaling the
limiting spectral radius in the complex Ginibre ensemble is also
Gumbel, with distribution function $e^{-e^{-t}}$. A similar result
holds in the quaternion case. The universality of the limiting
Gumbel law for spectral radius in any setting (real, complex, or
quaternion) has not been addressed.

The analog of Theorem \ref{mainthm} for complex Ginibre is a
triviality. The eigenvalues of the complex Ginibre ensemble form the
canonical radially symmetric determinantal process on the complex
plane, and it is the case that the moduli of the points of such a
process are independent. Stated in this generality this fact can be
found in \citet{MR2552864}, Chapter~7, but had been observed
earlier by \citet{MR1148410} specifically for complex Ginibre. While a Gumbel
law is the only possible scaling limit for the extremal point, the
precise scalings were worked out in \citet{MR1986426} where the author was
unaware of Kostlan's result, but rediscovered and used a~consequence
thereof. In the quaternion case there is nothing like this
determinantal trick. Still the expectations of certain eigenvalue
class functions factor nicely, which turns out to be enough; see
again \citet{MR1986426}.

For real Ginibre, there appear to be no shortcuts toward
pinning down the fluctuations of the spectral radius. Instead we
return to the standard random matrix theory machinery of tracing
through the limiting Fredholm determinant/Pfaffian formulas for the
related gap probabilities (which are available once the correlation
functions are known). The proof of Theorem \ref{mainthm} follows from
determining the real and complex gaps separately:

%
%
\begin{teo}
\label{CompThm}
Let $z_1, \ldots, z_n$ be the eigenvalues of $M$, then
\[
\mathbb{P} \biggl( \max_{k\dvtx  z_k \in\mathbb{C}/ \mathbb{R}} |z_k| \le\sqrt{n} +
\sqrt{ \frac
{\gamma_n}{4} } + \frac{t}{\sqrt{4 \gamma_n}} \biggr) \rightarrow
e^{-(1/2) e^{-t}}
\]
for any $t \in\mathbb{R}$ as $n \rightarrow\infty$, where again
$\gamma_n = \log(n/ (2\pi(\log n)^2) )$.
\end{teo}

%
%
\begin{teo}
\label{RealThm}
Introduce the integral operator $T$ with kernel
%
\begin{equation}
\label{Tkernel} T(x,y) = \frac{1}{\pi} \int_0^{\infty}
e^{-(x+u)^2} e^{-(y+u)^2} \,du.
\end{equation}
Let $\chi$ be the indicator of $(t,\infty)$. Then, as $n \rightarrow
\infty$,
%
\begin{equation}
\label{reallaw} \mathbb{P} \Bigl( \max_{k\dvtx  z_k \in\mathbb{R}} z_k
\le\sqrt{n} + t \Bigr) \rightarrow\sqrt{ {\det} (I - T \chi) \Gamma_t},
\end{equation}
where $\Gamma_t$ is built as follows. Set $g(x) = \frac{1}{\sqrt{2
\pi
}} e^{-x^2/2}$, $G(x) = \int_{-\infty}^x g(y) \,dy$ and denote by
$R(\cdot, \cdot)$ the kernel of the resolvent operator $(I - T \chi)^{-1}$. Then
\[
\Gamma_t = (1 - a_t) \biggl( 1 - \frac{1}{2} \int
_{-\infty}^t R(x,t) \,dx \biggr) + \frac{1}{2} ( 1
- b_t) \int_{-\infty}^t (I - T
\chi)^{-1} g(x) \,dx
\]
for $a_t=\int_t^{\infty} G(x) (I - T \chi)^{-1} g(x) \,dx$ and $b_t = (I
- T \chi)^{-1} G(t)$.
\end{teo}

Theorem \ref{mainthm} then just recasts the result for
complex points; the largest real eigenvalue simply lives on a smaller
scale. (Obviously the largest negative real eigenvalue exhibits the
same limit law.) To make this explicit, we check here that the
probability of the largest point in absolute value being real tends to zero.
With that event denoted by $A$ and $c_n = 1/4 \sqrt{\gamma_n}$,
\begin{eqnarray*}
\mathbb{P}(A) & \le&\mathbb{P} \Bigl( A, \max_k |
\lambda_k| \ge n^{1/2} + c_n \Bigr) + \mathbb{P}
\Bigl( \max_k | \lambda_k| \le
n^{1/2}+c_n \Bigr)
\\
& \le&\mathbb{P} \Bigl( \max_{k\dvtx  \lambda_k \in\mathbb{R}} \lambda_k \ge
n^{1/2} + M \Bigr)
\\
&&{} + \mathbb{P} \Bigl( \max_{k\dvtx  \lambda_k \in\mathbb
{C}/\mathbb{R}} | \lambda_k |
\le n^{1/2} + \sqrt{\gamma_n/4} - M/ \sqrt{4
\gamma_n} \Bigr)
\end{eqnarray*}
for any large $M$ as $n \uparrow\infty$. And by choice of $M$, the
lim(sup) of the right-hand side can be made arbitrarily small by the
outcomes of
Theorems \ref{CompThm} and \ref{RealThm}. Note by definition the
right-hand side of (\ref{reallaw}) is a distribution function, and so
tends to zero as $t$, here $M$, tends to infinity. This can also be
seen directly from the simple fact that $T\chi$ goes to zero in trace
norm in the same parameter limit.

From a technical standpoint the above means that we never
have to consider the mixed real/complex correlations. The same calculation
behind Theorem \ref{CompThm} produces the full Poisson point process
surrounding the Gumbel
limit. Since the complex eigenvalues occur in conjugate pairs and
(again) the real eigenvalues are in sub-scaling
the relevant statement is as follows.

%
%
\begin{cor}
\label{compcor}
Rescale the eigenvalues $\{ z_k\}$ lying in the (strict) upper half
plane as in
$z_k' = (r_k', \theta_k')$ with $r_k' = \sqrt{ 4 \gamma_n }  (
|z_k| - \sqrt{n} - \sqrt{ \frac{\gamma_n}{4}}  ) $,
$\theta_k' = \arg(z_k)$. The resulting point process converges, in the
sense of finite-dimensional distributions, to the Poisson random
measure with intensity $ \frac{1}{2 \pi} e^{-r}$ on $(-\infty,
\infty)
\times(0, \pi)$.
\end{cor}

A convergence result along the lines of Theorem \ref{RealThm} for the
largest real point
was
presented in \citet{forrester-2007}. There the right-hand side of
(\ref{reallaw}) was left in terms of the Fredholm Pfaffian of a $2
\times2$ matrix operator. Here, besides rigorously establishing the
appropriate norm convergence, the factorization in terms of scalar
operators coincides with the initial form of $\beta=1$ Tracy--Widom
($\mathrm{TW}_1$) distribution function as originally found for GOE. Indeed,
the resulting structure of the above limit law is precisely the same
as that form of $\mathrm{TW}_1$: $T$ replaces the Airy kernel, and the Gaussian
density $g(\cdot)$ plays the role of the Airy function $\mathrm{Ai}(\cdot)$.
Unsurprisingly, our derivation of (\ref{reallaw}) follows
\citet{MR1385083} quite closely.

One point of interest is that $T$, like the Airy or related Bessel
operator, is product Hilbert--Schmidt, and so trace class, on any
positive half-line. (Additional properties of $T$ are needed to show
that the factors which make up $\Gamma_t$ are sensible.) Unlike the
Airy or Bessel cases, however, $T$ does not posses a
Christoffel--Darboux form
and so is not integrable in the sense of
\citet{MR1730498}, or at least not in this simple way. This presents at
least one
roadblock in obtaining a ``closed'' expression for (\ref{reallaw}), say
something in terms of a single special function like the Painlev\'e
formulation of the Tracy--Widom laws. Even a characterizing PDE or
system of ODEs
has eluded us.
The full large deviations of (\ref{reallaw})
are also open. The right tail, as $t \rightarrow\infty$, is easily
seen to have a Gaussian shape; this again was pointed out in
\citet{forrester-2007}. The determination of the left tail lies
deeper and will be pursued in a later paper. This seems worthwhile
given the separate interest in the limiting largest real point due to
its applications in the stability analysis of certain biological
systems [\citet{1972Natur238413M}].

The next section recalls what is needed of the real Ginibre correlation
functions and gap probabilities. The limit laws for the largest complex
and real points are derived separately in Sections~\ref{sec3} and~\ref{sec4}.
Section~\ref{sec5} reports on some numerical simulations and discusses
additional open questions.

\section{Determinants}\label{sec2}

The results of \citet{borodin-2008} lead to (Fredholm) determinantal
formulas for the relevant gap probabilities in the $n \times n$ real
Ginibre ensemble $M$. We state things in the particular form that we need.

%
%
\begin{prop}
The probability $\mathbb{P}_{\mathbb{C},n}(t)$ of there being no
complex eigenvalues
of $M$ of modulus greater than $t > 0$ is given by
%
\begin{equation}
\mathbb{P}_{\mathbb{C},n}(t)^2 = \det ( I - K_n \chi )
\end{equation}
in which $K_n$ is a $2 \times2$ operator defined on $L^2(\mathbb{C}_+)
\oplus
L^2(\mathbb{C}_+)$, $\mathbb{C}_+ = \{ z\dvtx  \operatorname{Im}(z) >
0 \}$, cut down by the
indicator function
$\chi= \chi_{ \{ |z| > t \}} $. In the standard notation,
%
\begin{equation}
\label{complexK} K_n = \lleft[ \matrix{S_n &
DS_n
\vspace*{3pt}\cr
-IS_n & S_n^{\mathsf{T}}} \rright]
\end{equation}
with the various operators defined most easily through their kernels
%
\begin{equation}
\label{complexS} S_n(z,w) = \frac{i e^{-(1/2)(z - \bar{w})^2}}{\sqrt{ 2\pi}} (\bar {w} -  z) \phi(z)
\phi(w) e^{- z \bar{w}} \mathfrak{e}_{n-2} (z \bar{w}),
\end{equation}
$ DS_n(z,w) = - i S_n(z, \bar{w})$ and $IS_n(z,w) = i S_n (\bar{z}, w)$.
The shorthand
\[
\phi(z) = \sqrt{\operatorname{erfc} \bigl( \sqrt{2} \bigl| \operatorname {Im}(z) \bigr|
\bigr)}, \qquad\mathfrak{e}_n(z) = \sum_{k=0}^n
\frac{z^k}{k!},
\]
is used, where as usual $\operatorname{erfc}(z) = \frac{2}{\sqrt{\pi
}} \int_z^{\infty}e^{-t^2} \,dt$.
\end{prop}

Note that while \citet{borodin-2008} considers only even values of $n$
($n = 2M$ there), the subsequent results of \citet{sinclair-2008} shows
that the formulas remain unchanged for $n$ odd. This can be loosely
understood by considering that any ``extra'' particle must be real.

For gaps on the real line we have the following.

%
%
\begin{prop}\label{prop1}
The probability of there being no real eigenvalues of $M$ larger than
$t > -\infty$ is given by
\[
\mathbb{P}_{\mathbb{R},n}(t)^2 = \det ( I - K_n \chi ),
\]
where again $K_n$ is a $2 \times2$ operator now defined on
$L^2(\mathbb{R})
\oplus L^2(\mathbb{R})$,
and $\chi= \chi_{[t, \infty)}$. The overall form of $K_n$ is similar
to the complex case
%
\begin{equation}
\label{realK} K_n = \lleft[ \matrix{S_n & D
S_n
\vspace*{3pt}\cr
-IS_n + \epsilon& S_n^{\mathsf{T}}}
\rright].
\end{equation}
Here $\epsilon$ is the operator
%
\begin{equation}
\label{epsilonop} \epsilon f (x) = \frac{1}{2} \int_{-\infty}^{\infty}
\operatorname {sgn}(y-x) f(y) \,dy
\end{equation}
and the basic kernel is
%
\begin{eqnarray}\label{realS}
S_n(x,y) &=& \frac{e^{-(1/2) (x-y)^2}}{\sqrt{2 \pi}} e^{-xy}
\mathfrak {e}_{n-2}(xy)
\nonumber\\[-8pt]\\[-8pt]
&&{} + \frac{x^{n-1}e^{-(1/2) x^2}}{\sqrt{2 \pi} (n-2)!} \int_0^{y}
u^{n-2} e^{ -(1/2) u^2} \,du,\nonumber
\end{eqnarray}
in terms of which
$DS_n = \delta S_n^{\mathsf{T}}$ where
$\delta$ acts by differentiation on the first variable, and:
\begin{enumerate}
\item[(1)] when $n$ is even
\[
IS_n(x,y) = \epsilon S_n(x,y);
\]
\item[(2)] when $n$ is odd,
\begin{eqnarray*}
 IS_n(x,y) &=& \epsilon S_n(x,y) + \frac{1}{2^{n/2} \Gamma(n/2)}
\int_0^y u^{n-1} e^{-(1/2) u^2} \,du
\\
& =& \epsilon S_{n-1}(x,y)
\\
&&{} + \frac{1}{\sqrt{2 \pi} (n-2)!} \int _0^x\! \int_0^y
(w-u) (wu)^{n-2} e^{-(1/2) u^2 -(1/2) w^2} \,du \,dw.
\end{eqnarray*}
\end{enumerate}
In both cases $\epsilon$ also acts on the first variable.
\end{prop}

While the first form of the $IS_n$ kernel for $n$ odd may be more
attractive, the second is better for asymptotics.
Both structures are also valid for, for instance, for odd $n$ GOE. In
particular, in the first form the correction term
may be expressed as $\frac{\pi_{n-1}}{s_{n-1}}$ for $\pi_{n-1}$ the
relevant skew-orthogonal polynomial and $s_{n-1}$ its normalizer
(see below), and the same holds for GOE
with
appropriate substitutions for $\pi_{n-1}$ and $s_{n-1}$.

Note that since the $\epsilon$ operator is not trace class, what is
meant by the Fredholm determinant of (\ref{realK}) is at first not
clear. However, this is a standard technicality in the world of
$\beta=1$ ensembles, and how to mollify things or smooth out the
$\epsilon$ is well understood; see, for instance, \citet{MR1657844}, Section~VIII.

\begin{pf*}{Proof of Proposition~\ref{prop1}}
The $n$ even case is a restatement of \citet{borodin-2008}, Theorem~8. The $n$ odd case was stated without proof in
\citet{mays-forrester} (based on their derivation of the correlation
functions of $\beta=1$ ensembles of odd order); see also
\citet{sommers-wieczorek}. We include the odd
$n$ case here since the details of the
derivation have not before appeared in the literature. We
appeal to \citet{sinclair-2008}, Section~7 (providing yet another
derivation of the correlation kernel for $\beta=1$ ensembles of odd
order), using the particulars of Ginibre's real ensemble.

Recall the weighted skew-orthogonal polynomials for Ginibre's real
ensemble [\citet{forresternagao2008}] given by
\[
\pi_{2j}(x):= e^{-x^2/2} x^{2j}; \qquad
\pi_{2j+1}(x):= e^{-x^2/2} \bigl(x^{2j+1} - 2j
x^{2j-1} \bigr)
\]
with the normalization
\[
r_j:= \langle\pi_{2j}, \pi_{2j+1} \rangle= 2
\sqrt{2 \pi} (2j)!.
\]
We will not explicitly use the skew-inner product, but the interested
reader is referred to \citet{sinclair-2007}. One further set of
normalizations (which does not arise for even $n$) is
\[
s_{2j}:= \int_{-\infty}^{\infty}
\pi_{2j}(x) \,dx = 2^{j +
1/2} \Gamma(j + 1/2).
\]
(In general, similar normalizations are necessary for the odd degree
skew-orthogonal polynomials, but in this case these vanish.) In
particular,
%
\begin{equation}
\label{eq1} \frac{s_{2j}}{r_j} = \frac{2^{j+1/2} \Gamma(j+1/2)}{2 \sqrt{2 \pi}
(2j)!} = \frac{1}{2^{j+1} j!},
\end{equation}
where the last identity uses the Gamma function duplication formula.
We will also need
\[
\epsilon\pi_{2j+1}(x) = e^{-x^2/2} x^{2j}\quad
\mbox{and}\quad \epsilon\pi_{n-1}(x) = -2^{n/2-1}
\operatorname{sgn}(x) \gamma \biggl(\frac{n}2, \frac{x^2}2
\biggr).
\]

It suffices to establish (\ref{realS}) for $n$ odd, since the $DS_n$
and (the modified) $IS_n$ terms follow immediately from
\citet{sinclair-2008}. For $S_n$, \citeauthor{sinclair-2008} [(\citeyear{sinclair-2008}), page 31] implies
\begin{eqnarray*}
S_n(x,y) &= & S_{n-1}(x,y) - 2 \frac{\pi_{n-1}(x)}{s_{n-1}} \sum
_{j=0}^{J-1} \frac{s_{2j}}{r_j} \epsilon
\pi_{2j+1}(y)
\\
&&{}+ 2 \frac{\epsilon\pi_{n-1}(y)}{s_{n-1}} \sum_{j=0}^{J-1}
\frac{s_{2j}}{r_j} \pi_{2j+1}(x) + \frac{\pi_{n-1}(x)}{s_{n-1}}.
\end{eqnarray*}
Here, we use
%
\begin{equation}
\label{eq4} 2 \sum_{j=0}^{J-1}
\frac{s_{2j}}{r_j} \epsilon\pi_{2j+1}(y) = e^{-y^2/2} \sum
_{j=0}^{J-1} \frac{1}{2^j j!} y^{2j} =
e^{-y^2/2} \mathfrak e_{J-1} \bigl(y^2/2 \bigr)
\end{equation}
and
%
\begin{equation}
\label{eq5}
\frac{\epsilon\pi_{n-1}(y)}{s_{n-1}} = -\frac{\operatorname{sgn}(y)}{2} \frac{\gamma (n/2,
y^2/2 )}{\Gamma (n/2 )}
\end{equation}
to write
\begin{eqnarray*}
S_n(x,y) &=& S_{n-1}(x,y) + \frac{e^{-x^2/2} x^{n-1}}{2^{n/2}
\Gamma (n/2)} \bigl( 1 -
e^{-y^2/2} \mathfrak e_{J-1} \bigl(y^2/2 \bigr) \bigr)
\\
&&{} - \frac{ 2^{(n/2)-1}}{\sqrt{2\pi} (n-2)!} e^{-x^2/2} x^{n-2}
\operatorname{sgn}(y) \gamma \biggl(\frac{n}2, \frac{y^2}2
\biggr).
\end{eqnarray*}
Next, note that
\[
1 - e^{x^2/2} \mathfrak{e}_{J-1} \bigl(x^2/2 \bigr)
= \frac{1}{2^{(n-3)/2}
\Gamma((n-1)/2)} \int_0^x
u^{n-2} e^{-(1/2)u^2} \,du.
\]
More simply, $\gamma(\frac{n}{2}, \frac{x^2}{2} ) = 2^{-(n/2)
+1} \int_0^{|x|} u^{n-1} e^{-(1/2) u^2}$, so that
\begin{eqnarray*}
S_n(x,y) &=& S_{n-1}(x,y) + \frac{x^{n-1} e^{-(1/2)x^2}}{\sqrt{2
\pi} (n-2)!} \int
_0^y u^{n-2} e^{-(1/2) u^2} \,du
\\
&&{}- \frac{x^{n-2} e^{-(1/2) x^2}}{\sqrt{2 \pi} (n-2)!} \int_0^{y}
u^{n-1} e^{ -(1/2) u^2} \,du.
\end{eqnarray*}
Since $n-1$ is even we can substitute for $S_{n-1}$ to get
%
\begin{eqnarray}\label{SNodd}
S_n(x,y) &=& \frac{e^{-(1/2) (x-y)^2}}{\sqrt{2 \pi}} e^{-xy}
\mathfrak{e}_{n-3}(xy)\nonumber
\\
&&{} + \frac{x^{n-2}e^{-(1/2) x^2}}{\sqrt {2 \pi}
(n-3)!} \int_0^{y}u^{n-3} e^{ -(1/2) u^2} \,du
\nonumber\\[-8pt]\\[-8pt]
&&{}+ \frac{x^{n-1}
e^{-(1/2)x^2}}{\sqrt{2 \pi} (n-2)!} \int_0^y
u^{n-2} e^{ -(1/2) u^2} \,du\nonumber
\\
&&{}  - \frac{x^{n-2} e^{-(1/2) x^2}}{\sqrt{2 \pi} (n-2)!} \int
_0^{y} u^{n-1} e^{ -(1/2) u^2} \,du.\nonumber
\end{eqnarray}
Integration by parts on the second term yields
\begin{eqnarray*}
&&\frac{x^{n-2}e^{-(1/2) x^2}}{\sqrt{2 \pi}
(n-3)!} \int_0^{y}
u^{n-3} e^{ - (1/2) u^2} \,du
\\
&&\qquad = \frac{e^{-(1/2) x^2} e^{-(1/2) y^2} (xy)^{n-2}}{\sqrt{2\pi}
(n-2)!} + \frac{x^{n-2} e^{-(1/2) x^2}}{\sqrt{2\pi} (n-2)!} \int_0^y
u^{n-1} e^{-(1/2) u^2} \,du.
\end{eqnarray*}
The first of these terms takes
\[
\frac{e^{-(1/2) (x-y)^2}}{\sqrt{2 \pi}} e^{-xy} \mathfrak{e}_{n-3}(xy) \quad\mbox{to}\quad\frac{e^{-(1/2)
(x-y)^2}}{\sqrt{2 \pi}} e^{-xy} \mathfrak{e}_{n-2}(xy)
\]
and the second cancels the last term in (\ref{SNodd}) to produce the
advertised formula.
\end{pf*}

\subsection{From correlations to gap probabilities}\label{sec2.1}

Once again, what \citet{borodin-2008} and \citet{sinclair-2008} establish
are Pfaffian formulas for the $k$-point (any combination of real and
complex) correlation functions.
For completeness we briefly review how to go from the correlations to
the above determinantal gap formulas.

The situation is similar to the
classical $\beta= 1$ (GOE) situation, but is complicated by the
presence of both real and complex eigenvalues. Letting $L$ represent
the number of real eigenvalues and $M$ the number of complex conjugate
eigenvalues, there is a different joint eigenvalue density for each
pair $(L,M)$ with $L + 2M = n$. Representing this density as
$\Omega_{L,M}\dvtx  \mathbb{R}^L \times\mathbb{C}^M \rightarrow[0,
\infty)$ [the exact
formula for which can be found in \citet{1991PhRvL67941L}, \citet{MR1437734}], the
normalization constant for the ensemble is given by
\[
Z_n = \mathop{\sum_{(L, M)}}_{L + 2M = n} \frac{1}{L! M! 2^M}
\int_{\mathbb{R}^L} \int_{\mathbb{C}^M}
\Omega_{L,M}(\bolds\alpha, \bolds\beta) \,d\mu_{\mathbb{R}
}^L(
\bolds \alpha) \,d\mu_{\mathbb{C}}^M(\bolds\beta),
\]
where $\mu_{\mathbb{R}}$ and $\mu_{\mathbb{R}}^L$ are Lebesgue
measure on $\mathbb{R}$ and
$\mathbb{R}^L$, and $\mu_{\mathbb{C}}$ and $\mu_{\mathbb{C}}^M$
are defined analogously.\vadjust{\goodbreak}

We define the $\ell,m$ correlation function $R_{\ell,m}\dvtx
\mathbb{R}^{\ell} \times\mathbb{C}^m \rightarrow[0,\infty)$ by
\begin{eqnarray*}
&& R_{\ell, m}(\mathbf x, \mathbf z)
\\
&&\qquad = \frac{1}{Z_n} \mathop{\sum
_{(L,M)}}_{L
\geq\ell, M \geq m} \frac{1}{(L-\ell)! (M - m)! 2^{M-m}}
\\[-2pt]
&&\hspace*{88pt}{}\times\int_{\mathbb{R}^{L-\ell}} \int_{\mathbb{C}^{M-m}} \Omega
_{L,M}(\mathbf x \vee\bolds \alpha, \mathbf z \vee\bolds\beta) \,d
\mu_{\mathbb{R}}^{L-\ell
}(\bolds\alpha ) \,d\mu_{\mathbb{C}}^{M-m}(
\bolds\beta),
\end{eqnarray*}
where, for instance, $\mathbf x \vee\bolds\alpha= (x_1,\ldots,x_{\ell},
\alpha_1, \ldots, \alpha_{L-\ell})$. That is, the $\ell,m$
correlation function is a weighted sum of all marginal densities
formed by integrating out $L - \ell$ real variables and $M - m$
complex variables from all $\Omega_{L,M}$ for which this makes sense.

The main result of \citet{borodin-2008} demonstrates the existence of
three matrix
kernels
$K_{n}^{\mathbb{R},\mathbb{R}}, K_n^{\mathbb{C}, \mathbb{C}}$ and
$K_n^{\mathbb{R}, \mathbb{C}}$ (and its
transpose $K_n^{\mathbb{C}, \mathbb{R}}$) such that
\[
R_{\ell, m}(\mathbf x, \mathbf z) = \operatorname{Pf}\lleft[\matrix{
\bigl[ K_{n}^{\mathbb{R}, \mathbb{R}}(x_j, x_k)
\bigr]_{j,k=1}^{\ell\times\ell} & \bigl[ K_{n}^{\mathbb{R}, \mathbb{C}}(x_j,
z_t) \bigr]_{j,t=1}^{\ell
\times m}
\vspace*{4pt}\cr
\bigl[
K_{n}^{\mathbb{C}, \mathbb{R}}(z_s, x_k)
\bigr]_{s, j=1}^{m \times
\ell} & \bigl[ K_{n}^{\mathbb{C}, \mathbb{C}}(z_s,
z_t) \bigr]_{s,t=1}^{m
\times m} } \rright].
\]
Now consider a $C \subseteq\mathbb{C}$ which is
invariant under complex conjugation, written as the
disjoint union $C = A \cup B$ where $A \subseteq\mathbb{R}$ and $B
\subseteq
\mathbb{C}\setminus\mathbb{R}$. Conditioning on the number of
real and complex
conjugate pairs of eigenvalues, we have that
\begin{eqnarray*}
\mathbb{P}_{C, n} &:=& P (\mbox{no eigenvalues in }C)
\\
& =& \mathop{\sum_{(L,M)}}_{L + 2M = n} P \bigl( \mbox{exactly $L$ real
eigenvalues and all eigenvalues in $C^c$} \bigr)
\\
& =& \frac{1}{Z_n} \mathop{\sum_{(L,M)}}_{L + 2M = n}
\frac{1}{L! M!
2^M}\int_{\mathbb{R}^L}\! \int_{\mathbb{C}^M}
\Biggl\{ \prod_{j=1}^L \bigl(1 -
\chi_A(\alpha_j) \bigr)\prod
_{k=1}^M \bigl(1 - \chi_B(\beta
_k) \bigr) \Biggr\}
\\
&&\hspace*{130pt}{}\times\Omega_{L, M}(\bolds\alpha, \bolds\beta) \,d
\mu_{\mathbb{R}}^L(\bolds\alpha) \,d\mu_{\mathbb{C}}^M(
\bolds \beta)
\end{eqnarray*}
with $\chi_A$ and $\chi_B$ the
characteristic functions of $A$ and $B$.
Expanding the products in the integrand and simplifying leads to
\[
\mathbb{P}_{C,n} = \mathop{\sum_{(\ell, m)}}_{\ell+ 2m \leq n}
\frac
{(-1)^{\ell
+ m}}{\ell! m! 2^m} \int_{A^{\ell}}\! \int_{B^m}
R_{\ell, m}(\mathbf x, \mathbf z) \,d\mu_{\mathbb{R}}^{\ell}(
\mathbf x) \,d\mu_{\mathbb
{C}}^m(\mathbf z).
\]
In particular, when $B = \varnothing$, that is, when we are
interested in the probability that there are no eigenvalues in some
subset $A$ of $\mathbb{R}$, but we place no restrictions on the complex
eigenvalues,
\begin{eqnarray*}
\mathbb{P}_{A, n} &=& \sum_{\ell=1}^n
\frac{(-1)^{\ell}}{\ell!} \int_{A^{\ell}} R_{\ell,
0}(\mathbf x, -)
\,d\mu_{\mathbb{R}}^{\ell}(\mathbf x)
\\
&=& \sum_{\ell=1}^n \frac{(-1)^{\ell}}{\ell!} \int
_{A^{\ell}} \operatorname{Pf} \bigl[ K_{n}^{\mathbb{R},
\mathbb{R}}(x_j,
x_k) \bigr]_{j,k=1}^{\ell\times\ell} \,d\mu_{\mathbb{R}}^{\ell
}(
\mathbf x).
\end{eqnarray*}
Similarly, if $A = \varnothing$,
\begin{eqnarray*}
\mathbb{P}_{B, n} &=& \sum_{m=1}^{n/2}
\frac{(-1)^{m}}{m!} \int_{B^m} \operatorname{Pf} \biggl[
\frac{1}2 K_{n}^{\mathbb{C}, \mathbb{C}
}(z_s,
z_t) \biggr]_{s,t=1}^{m \times m} \,d\mu_{\mathbb{C}}^{m}(
\mathbf z)
\\
&=& \sum_{m=1}^{n/2} \frac{(-1)^{m}}{m!} \int
_{(B^+)^m} \operatorname{Pf} \bigl[K_{n}^{\mathbb{C}, \mathbb{C}}(z_s,
z_t) \bigr]_{s,t=1}^{m \times m} \,d\mu_{\mathbb{C}}^{m}(
\mathbf z),
\end{eqnarray*}
where $B^+$ is the component of $B$ which lies in the upper half
plane. Each of the last two displayed equations defines the Fredholm
Pfaffian of the indicated matrix kernel $K_n$, or, in symbols
$\operatorname{Pf} (J - K)$ where $J =
\bigl[{\fontsize{8.36}{8}\selectfont{\matrix{ 0 &\!\! 1 \cr-1 &\!\! 0}}} \bigr] \otimes I$, and $I$ is the
\mbox{$n\times n$} identity matrix. One may then invoke
the relationship between the Fredholm Pfaffian and the Fredholm
determinant [\citet{rains-2000}, \citet{borodin-2007-40}],
\[
\det(I + J K) = \operatorname{Pf}(J - K)^2.
\]

\section{Complex points}\label{sec3}

Returning to (\ref{complexK}) and (\ref{complexS}) the correct scaling
can be implemented as in $z, w \mapsto Z_n, W_n$,
%
\begin{eqnarray}\label{scaling1}
Z_n(r, \theta) &=& \biggl(\sqrt{n} + \sqrt{
\frac{\gamma_n}{4} } + \frac
{r}{\sqrt{4 \gamma_n}} \biggr) e^{i \theta},
\nonumber\\[-8pt]\\[-8pt]
W_n(s, \eta) &=& \biggl(\sqrt {n} + \sqrt{ \frac{\gamma_n}{4} } +
\frac{s}{\sqrt{4 \gamma_n}} \biggr) e^{i\eta},\nonumber
\end{eqnarray}
where $\gamma_n = \log(n/ (2\pi(\log n)^2 ))$, and it is implicit that
$r, s > -\gamma_n$. Next, we replace $K_n(z, w)$ by
%
\begin{equation}
\label{scaling2} \widetilde{K}_n (r, \theta; s, \eta) = \sqrt{
\frac{ |Z_n| |W_n|
}{ 4
\gamma
_n}} K_n (Z_n, W_n),
\end{equation}
the variable change occurring entry-wise in $\widetilde{K}_n$, which
acts on
\[
\mathcal{L}_t = L^2(\mathcal{T}) \oplus L^2(
\mathcal{T}), \qquad \mathcal{T} = [t, \infty) \times(0, \pi) \ni(r, \theta), (s,
\eta) \qquad t > -\infty.
\]
The restriction to $\mathcal{T}$ is from now on assumed in the
definition of the transformed~$\widetilde{K}_n$.

The typical procedure (which is followed in the real case) is to
identify a limit kernel/operator $K$ on $\mathcal{T}$ for which
$\widetilde{K}_n \rightarrow K$ in trace norm, concluding the
convergence of
$\det(I -K_n)$. Here instead, though $\widetilde{K}_n$ is trace class
(it is
finite rank), it is more convenient to cast things in the
Hilbert--Schmidt norm, in which $\widetilde{K}_n$ vanishes in the limit.
This prompts the introduction of the regularized determinant
\[
{\det}_2 (I + A) = \det \bigl( (I + A) e^{-A} \bigr).
\]
In particular, if $A$ is Hilbert--Schmidt with eigenvalues $\{
\lambda_k(A) \}_{k \ge0}$, there is the evaluation $ {\det}_2 (I +
A) = \prod_{k=0}^{\infty} (1 + \lambda_k) e^{-\lambda_k}$
[\citet{MR1744872}, Section~IV.7] allowing us to write
\[
\det(I - \widetilde{K}_n ) = {\det}_2(I -
\widetilde{K}_n ) e^{-\operatorname{tr}\widetilde{K}_n}.
\]
With a matrix kernel, the trace is just the sum of the traces of the
diagonal entries.
The proof of Theorem \ref{CompThm} is then completed via the basic
estimate, with $\| \cdot\|$
the Hilbert--Schmidt norm,
%
\begin{equation}
\label{HSdif} \qquad \bigl| {\det}_{2} (I + A) - {\det}_{2} (I+B) \bigr|
\le\| A -B \| \exp \bigl( \tfrac{1}{2} \bigl(\|A\| + \|B\| +1\bigr)^2
\bigr);
\end{equation}
again see \citet{MR1744872}, Section~IV.7, along with the next lemma.

%
%
\begin{lemma}
\label{NormConvergence}
We have that $\| \widetilde{S}_n \|_2 \rightarrow0$ as $n \rightarrow
\infty
$ while
\[
\operatorname{tr}( \widetilde{S}_n) \rightarrow\tfrac{1}{2}
e^{-t}.
\]
In addition, $ \| \widetilde{DS}_n \|_2 = \| \widetilde{IS}_n \|_2
\rightarrow0$.
All norms are with respect to $L^2(\mathcal{T} \times\mathcal{T})$.
\end{lemma}

In particular, estimate (\ref{HSdif}) gives the desired result upon
choosing $A= \widetilde{K}_n$ and $B=0$. For a different way to understand
Theorem \ref{CompThm}, the proof of Lemma \ref{NormConvergence} will
show that on bounded sets, the kernel $\widetilde{S}_n$ is well approximated
(after an unimportant conjugation) by
%
\begin{equation}
\label{Skappa} {S}_{\kappa}(r, \theta; s, \eta) = \frac{\kappa}{2 \pi}
\frac{ e^{-(1/2)( r+s)}}{ (1 + \kappa) e^{i (\theta- \eta)} -1},
\end{equation}
in which $\kappa= \kappa(n)$ tends to zero as $n \rightarrow\infty$.
[So, formally, the limit operator has kernel $\frac{1}{2 \pi} \chi
_{\{
\theta\}}(\eta) e^{-(1/2)( r+s)}$. Again though, we do not attempt
to carry out a proof in this manner.]
Here the decoupling of the moduli and phases of the points is made
explicit, as is the asymptotic independence\vspace*{1pt} of neighboring phases (on
the scale $\kappa$).
The kernels for $\widetilde{DS}_n$ and $\widetilde{IS}_n$ will be
shown to exhibit a
shaper decay. One can also check that $\det(I - S_{\kappa}) \sim
e^{-(1/2) e^{-t}}$ from the series definition of the Fredholm
determinant.

As for the proof of Lemma \ref{NormConvergence}, we will use the next
three estimates on the polynomial $\mathfrak{e}_n(z)$.

%
%
\begin{lemma}[{[Wimp-\citet{MR2288197}]}]
\label{wimp}
Uniformly in $t \ge0$,
\[
e^{-n t} \mathfrak{e}_n(nt) = \mathbh{1}_{0 \le t < 1} +
\frac
{1}{\sqrt{2}} \frac{\mu(t) t}{ t-1} \operatorname{erfc} \bigl( \sqrt{n} \mu(t)
\bigr) \biggl( 1 + O \biggl( \frac
{1}{\sqrt{n}} \biggr) \biggr),
\]
where $\mu(t) = \sqrt{t - \log t - 1}$ is taken positive for all $t$.
\end{lemma}

%
%
\begin{lemma}[{[\citet{MR2264712}]}]\label{LBM}
For small enough $\delta> 0$, let now $\mu(z) = \sqrt{z - \log z -1}$
be uniquely defined as analytic in $|z-1| < \delta$
with $\mu(1+x) > 0$ for $0 < x< \delta$. Then, for any $M > 1$ it holds
that with $n \rightarrow\infty$,
\[
e^{-nz} \mathfrak{e}_n(nz) = \frac{1}{2 \sqrt{2} \mu'(z)}
\operatorname{erfc} \bigl(\sqrt{n} \mu (z) \bigr) \biggl( 1 + O \biggl(
\frac{1}{n(z-1)} \biggr) \biggr)
\]
for $z$ satisfying $\frac{M}{\sqrt{n}} \le|z-1| \le\delta$ and
$|{\arg(z-1)}| \le\frac{2 \pi}{3}$.
\end{lemma}

%
%
\begin{lemma}[{[\citet{MR2424215}]}]\label{LKKMM}
For any $0 < \alpha< 1/2$, set $U = \{ |z-1| \le n^{-\alpha}\}$, and denote
by ${D}$ the unit disk. Then
\[
\mathfrak{e}_{n-1}(nz) = e^n z^n
\frac{1}{\sqrt{2 \pi n} (1-z)} \biggl( 1 + O \biggl( \frac{1}{n | 1-z|^2} \biggr) \biggr)\qquad\mbox{for } z \in {D}^c - U
\]
and all $n > 1$.
\end{lemma}

Both \citet{MR2264712} and \citet{MR2424215} contain more detailed and
complete asymptotics along the lines stated in Lemmas \ref{LBM} and
\ref{LKKMM}; we record only what is used here. Also, as is easy to
check, the appraisals of all three lemmas apply without change to
$\mathfrak{e}_{n-2}$ (rather than say $\mathfrak{e}_n, \mathfrak
{e}_{n-1}$) for $n$
large enough. Last we should point out that Lemma \ref{wimp} may be
arrived at by combining Lemmas \ref{LBM} and \ref{LKKMM} (the
asymptotics must be consistent after all). It is convenient though to
have a single result to quote for the full range of the real
argument.

\begin{pf*}{Proof of Proposition \ref{NormConvergence}}
This is split
into five steps. Throughout, $C$ is a large positive constant that may
change from one line to the next.

\textit{Step} 1 is the trace calculation, integrating
\begin{eqnarray*}
\widetilde{S}_n (r,\theta; r, \theta) & =& \frac{|Z_n|}{2 \sqrt {\gamma_n}}
S_n(Z_n, Z_n)
\\
& =& \frac{1}{\sqrt{2 \pi}} \frac{|Z_n|}{\sqrt{\gamma_n}} \operatorname{Im} {(Z_n)}
e^{2 \operatorname{Im}(Z_n)^2} \operatorname{erfc} \bigl(\sqrt{2} \operatorname{Im}(Z_n)
\bigr) e^{-|Z_n|^2} \mathfrak {e}_{n-2} \bigl(|Z_n|^2
\bigr)
\end{eqnarray*}
with again $Z_n = (\sqrt{n} + \sqrt{ \frac{\gamma_n}{4} } + \frac
{r}{\sqrt{4 \gamma_n}})e^{i\theta}$ over $(r, \theta) \in\mathcal{T}$,
recall (\ref{scaling1}), (\ref{scaling2}).

Since,\vspace*{-2pt} for real $y > 0$ we have that $ \operatorname{erfc}(y) \le
\frac{1}{\sqrt {\pi}
y } e^{-y^2}$ while $\operatorname{erfc}(y) = \frac{1}{\sqrt{\pi} y
} e^{-y^2}(1+
O(1/y^2))$ for $y \rightarrow\infty$,
%
\begin{equation}
\label{trace1} \lim_{ n \rightarrow\infty} \frac{1}{\sqrt{2 \pi}} \operatorname
{Im} {(Z_n)} e^{2 \operatorname{Im}
(Z_n)^2} \operatorname{erfc} \bigl(\sqrt{2}
\operatorname{Im}(Z_n) \bigr) = \frac{1}{2\pi},
\end{equation}
pointwise on $\mathcal{T}$, with the left-hand side being bounded by
the right for all large $n$.

Next, with
%
\begin{equation}
\label{tn} t_n = \frac{1}{n} |Z_n|^2
= 1 + \frac{\sqrt{\gamma_n} + r/\sqrt {\gamma
_n}}{\sqrt{n}} + \frac{( \sqrt{\gamma_n} + r/\sqrt{\gamma_n})^2}{4n},
\end{equation}
Lemma \ref{wimp} implies that
%
\begin{equation}
\label{diag}\quad  e^{-|Z_n|^2} \mathfrak{e}_{n-2}
\bigl(|Z_n|^2 \bigr) = \frac{1}{\sqrt{2}}
\frac
{\mu(t_n)
t_n}{ t_n-1} \operatorname{erfc} \bigl( \sqrt{n} \mu(t_n) \bigr)
\biggl( 1 + O \biggl( \frac
{1}{\sqrt{n}} \biggr) \biggr)
\end{equation}
uniformly on $\mathcal{T}$. And since $\mu(1 + \varepsilon) = \frac
{\varepsilon}{\sqrt{2}}( 1+ O (\varepsilon))$ for $ 0 \le
\varepsilon
\ll1$, (\ref{tn})\break and~(\ref{diag}) produce
%
\begin{equation}
\label{trace2}\quad
\frac{|Z_n|}{\sqrt{\gamma_n}} e^{-|Z_n|^2} \mathfrak {e}_{n-2}
\bigl(|Z_n|^2 \bigr) = e^{-r} \bigl(1+o(1)
\bigr)\qquad\mbox{uniformly for } r = o(\sqrt{\gamma_n}).\hspace*{-25pt}
\end{equation}
Here we have used that
%
\begin{equation}
\label{scalingdef} \frac{\sqrt{n}}{\sqrt{2 \pi} \gamma_n} e^{-\gamma_n/2} \rightarrow1,
\end{equation}
which in effect dictates the choice of $\gamma_n$. For the tail we have
the following bounds: with $r \ge0$ and $n$ large enough so that
$\gamma_n \ge1$ while $\gamma_n^{3/2}/n^{1/2} \le1/2$,
\begin{eqnarray*}
\frac{|Z_n|}{\sqrt{\gamma_n}} e^{-|Z_n|^2} \mathfrak {e}_{n-2}
\bigl(|Z_n|^2 \bigr) & \le& \frac{\sqrt{n}}{\gamma_n}
t_n^2 \exp \bigl( - n \mu^2(t_n)
\bigr)
\\
& \le& C \frac{\sqrt{n}}{\gamma_n} r^2 \exp \biggl( - \frac{n}{2}
\biggl( \sqrt{\frac{\gamma_n}{n}} - \frac{\gamma_n}{n} \biggr) \biggl(
\frac{\sqrt{\gamma_n} + r/\sqrt{\gamma_n}}{\sqrt{n}} \biggr) \biggr)
\\
& \le& C r^2 e^{- r/4}.
\end{eqnarray*}
The second line uses the inequality
%
\begin{equation}
\label{littlebound} \varepsilon- \log(1+\varepsilon) \ge\tfrac{1}{2} \bigl(
\delta- \delta^2 \bigr) \varepsilon\qquad\mbox{for all } \varepsilon
\ge\delta\mbox{ and } 0 \le\delta< 1
\end{equation}
with the choices $\varepsilon= t_n -1$ and $\delta= \sqrt{\gamma
_n/n}$. Hence, dominated convergence coupled with (\ref{trace1}) and
(\ref{trace2}) yields
\[
\operatorname{tr}(\widetilde{S}_n) = \int_{\mathcal{T}}
\widetilde {S}_n (r,\theta; r, \theta ) \,dr \,d \theta= \frac{1}{2}
e^{-t} \bigl(1+o(1) \bigr)
\]
as required.\vadjust{\goodbreak}

\textit{Step} 2 considers $\widetilde{S}_n$ away (though just barely) from
the diagonal.
All further nontrivial behavior occurs when the argument of $\mathfrak
{e}_{n-2}(Z_n \widebar{W}_n)$ is in a small neighborhood of $ \frac
{1}{n} Z_n \widebar{W}_n =1$ for which we can invoke Lemma \ref{LBM}.
For given $\theta\in(0, \pi)$ consider the set
\[
\mathcal{N}_{\theta, n} = \biggl\{ (\eta, r, s)\dvtx  \eta\in(0, \pi), | \theta -
\eta| \le\frac{1}{n^{1/4} \sqrt{\gamma_n}}, t \le r, s \le{ n }^{1/4} \biggr\} \cap
\mathcal{T},
\]
in connection to which it will be useful to make the definition
%
\begin{equation}
\label{epschoice} \varepsilon= \frac{1}{n^{1/4} \sqrt{ \gamma_n} }.
\end{equation}
Similar to before, define
$z_n = \frac{1}{\sqrt{n}} Z_n$, $w_n = \frac{1}{\sqrt{n}} W_n$, and record
%
\begin{eqnarray}\label{tau}
z_n \bar{w}_n &=& \biggl( 1 + \sqrt{\frac{\gamma_n}{n}} + \frac{r+s}{2
\sqrt{\gamma_n n}}
\nonumber\\[-8pt]\\[-8pt]
&&\hspace*{5pt}{} + \frac{ ( \sqrt{\gamma_n} + r/\sqrt{\gamma_n}) ( \sqrt{\gamma_n} +
s/\sqrt{\gamma_n})}{4n} \biggr)
e^{-i (\theta-\eta)}.\nonumber
\end{eqnarray}
Note $|z_n|-1, |w_n|- 1 = O(\varepsilon)$ on $ \mathcal{N}_{\theta, n}
$, and that $z _n \bar{w}_n$ satisfy the assumptions of Lemma \ref{LBM} there,
\[
\frac{1}{C} \sqrt{\frac{\gamma_n}{n}} \le|1- z_n
\bar{w}_n| \le C \varepsilon,\qquad\bigl| \arg(1 - z_n
\bar{w}_n)\bigr| \le\pi/2.
\]
As we still have the estimate $\operatorname{erfc}(z) = \frac
{e^{-z^2}}{\sqrt{\pi} z}
(1 + O(1/z^2))$ for $\arg(z) < 3\pi/4$, it holds that
%
\begin{equation}
\label{mainoffdiag}
e^{-Z_n \widebar{W}_n } \mathfrak{e}_{n-2}( Z_n
\widebar{W}_n) = \frac
{1}{\sqrt{2 \pi n } } \frac{z_n \bar{w}_n}{ z_n \bar{w}_n -1}
e^{-n
\mu
^2 (z_n \bar{w}_n)} \biggl(1+ O \biggl( \frac{1}{\sqrt{n \gamma_n}} \biggr) \biggr),\hspace*{-30pt}
\end{equation}
since $\mu'(z) = \frac{z-1}{2 z \mu(z)}$. The rational term in (\ref
{mainoffdiag}) may be bounded roughly as
%
\begin{equation}
\label{polyfactorbound} \biggl\llvert \frac{z_n \bar{w}_n}{ z_n \bar{w}_n -1} \biggr\rrvert \le C \sqrt {
\frac{n}{\gamma_n}}\qquad\mbox{on } \mathcal{N}_{\theta, n}.
\end{equation}
And combining the exponent in (\ref{mainoffdiag}) with $e^{-(1/2)
(Z_n - \widebar{W}_n)^2} \phi(Z_n) \phi(W_n)$, $S_n$ has the overall
exponential factor of $n$ times
%
\begin{eqnarray}
\label{mainexpexpand}
\qquad && - \tfrac{1}{2} (z_n - \bar{w}_n)^2
-  \operatorname{Im}(z_n)^2 - \operatorname{Im}(w_n)^2
- \mu(z_n \bar{w}_n)^2\nonumber
\\
&&\qquad = 1 - \tfrac{1}{2} \bigl(|z_n|^2 +|w_n|^2 \bigr) + \log\bigl( |z_n||w_n| \bigr)
\\
&&\quad\qquad{}  + i \bigl( \bigl( \operatorname{Re}(z_n) \operatorname{Im}(z_n)
+ \arg (z_n) \bigr) - \bigl(\operatorname{Re}(w_n)
\operatorname{Im} (w_n) + \arg(w_n) \bigr) \bigr),\nonumber
\end{eqnarray}
where we are assuming $\operatorname{Im}(z_n), \operatorname{Im}(w_n) >0$.
The relevant part of (\ref{mainexpexpand}) satisfies
%
\begin{eqnarray}\label{mainexpbound}
&& 1 - \frac{1}{2} \bigl(|z_n|^2 +
|w_n|^2 \bigr) + \log\bigl( |z_n|
|w_n| \bigr)\nonumber
\\
&&\qquad \le  - \frac{1}{2} \bigl[\bigl(|z_n|-1\bigr)^2+ \bigl(|w_n|^2 -1 \bigr)^2 \bigr]
\nonumber\\[-8pt]\\[-8pt]
&&\quad\qquad{} - \biggl( \frac{1}{2} \bigl(|z_n| + |w_n|\bigr) - 1\biggr)
\biggl( \frac{1}{2} \sqrt{\frac{\gamma_n}{n} } - \frac{\gamma_n}{4n}
\biggr)\nonumber
\\
&&\qquad \le - \frac{\gamma_n}{2 n} - \frac{r+s}{8n} - \frac{ r^2 + s^2}{8
\gamma_n n} +\frac{\gamma_n^{3/2}}{n^{3/2}}\nonumber
\end{eqnarray}
for all $r, s\ge0$ and $\gamma_n/n \le1$.
Here again (\ref{littlebound}) is used (twice) with \mbox{$\varepsilon=
|z_n|-1$}, $|s_n| -1$ and $\delta= \frac{1}{2} \sqrt{\frac{\gamma_n}{n}}$.
(For $r$, $s < 0$, or in particular just bounded, a Taylor expansion
produces a better bound, with $\frac{r+s}{8n}$ replaced by $\frac{r+s}{2}$.)

Finally, there remains the prefactor,
%
\begin{eqnarray}
\label{prefactor}
&& \frac{1}{4\pi} \sqrt{\frac{{n}}{{\gamma_n }}} \times \sqrt {
\frac{|z_n| |w_n| }{\operatorname{Im}(z_n) \operatorname{Im}(w_n)} } i (\bar{w}_n - z_n)
\nonumber\\[-8pt]\\[-8pt]
&&\qquad = \frac{1}{2\pi} \sqrt{\frac{{n}}{{\gamma_n }}} \times\frac{
\sin((\theta+ \eta)/2) + \delta_n}{\sqrt{\sin(\theta) \sin
({\eta})}}
e^{- (i/2) (\theta- \eta)},\nonumber
\end{eqnarray}
where $\delta_n$ is an additive error term that satisfies $\delta_n = O
(n^{-1/4} \gamma_n^{-1/2} ) = O(\varepsilon)$ for
$r, s \le{n}^{1/4}$.

Combining the above [and recalling (\ref{scalingdef})], we have the
upper bound
%
\begin{equation}
\label{FinalBound1} \bigl\llvert \widetilde{S}_n( r, \theta; s, \eta)
\bigr\rrvert \le C \frac{\sin((\theta+ \eta)/2) + \varepsilon}{\sqrt{\sin(\theta) \sin({\eta})}} e^{-(r+s)/8}\qquad\mbox{on }
\mathcal{N}_{\theta, n}.
\end{equation}
And since
%
\begin{equation}
\label{dumbtrig} \int_{\varepsilon}^{\pi/2}\!\! \int
_{\theta- \varepsilon}^{\theta+
\varepsilon} \frac{ (\sin((\theta+ \eta)/2) + \varepsilon)^2}{\sin
(\theta)
\sin({\eta})} \,d\eta \,d \theta \le
C \varepsilon 
\end{equation}
(this for any small $\varepsilon> 0$) with a similar bound in a
neighborhood of $\theta= \pi$,
it follows that
\[
\int_{\varepsilon}^{\pi- \varepsilon} \,d \theta\int_{\mathcal{N}_{n,
\theta}}
| \widetilde{S}_n |^2 \le C \varepsilon,
\]
where now recall that $\varepsilon= \varepsilon_n \downarrow0$.\vadjust{\goodbreak}

For the integral over $0 < \theta< \varepsilon$, we go back to the
start and
bound the $\operatorname{erfc}(a)$ appearing in each copy of $\phi
(a)$ in a simpler
way: for $a > 0$, $\operatorname{erfc}(a) = \frac{2}{\sqrt{\pi}}
\int_0^{\infty}
e^{-(t+a)^2} \,dt  \le e^{-a^2}$. This removes the integrability issues
due to the inverse sines in (\ref{FinalBound1}) at the expense of an
additional factor of $\sqrt{n}$. Importantly though this keeps the over
all exponent from (\ref{mainexpexpand}) unchanged. In particular, the estimate
\[
\bigl\llvert \widetilde{S}_n ( r, \theta; s, \eta) \bigr\rrvert \le
C \sqrt{n} \biggl( \sin \biggl( \frac{\theta+ \eta}{2} \biggr) + \varepsilon \biggr)
e^{-(r+s)/8},
\]
is also available on any $\mathcal{N}_{\theta, n}$. On the region of
current interest, $\sin ( \frac{\theta+ \eta}{2}  ) = O
(\varepsilon)$, and so
\[
\int_0^{\varepsilon}\! \int_{\mathcal{N}_{\theta, n}} |
\widetilde{S}_n|^2 \le C n \varepsilon^4,
\]
explaining in part the choice that $\varepsilon$ should decay a bit
faster than $n^{1/4}$.

%
%
\begin{rem*} If we further restrict $r,s = o (\sqrt{\gamma_n})$ the
bounds (\ref{polyfactorbound}) and (\ref{mainexpbound}) can be
supplanted by
\[
\frac{z_n \bar{w}_n}{ z_n \bar{w}_n -1} = \frac{e^{i (\theta- \eta)}}{
(1 + \sqrt{\gamma_n/n}) e^{i (\theta-\eta)} -1 } \bigl(1 + o(1) \bigr)
\]
and
\[
1 - \frac{1}{2} \bigl(|z_n|^2 +
|w_n|^2 \bigr) + \log\bigl( |z_n|
|w_n| \bigr) = - \frac
{\gamma_n}{2n} - \frac{r+s}{2n} + o \biggl(
\frac{1}{n} \biggr),
\]
respectively. Also considering a fixed $\theta\in(0, \pi)$ with
$|\theta- \eta| = o(1)$, one has $ \frac{ \sin((\theta+\eta)/2)}{\sqrt{\sin(\theta) \sin({\eta})}} = 1 + o(1)$.

Therefore, setting $f_n(r, \theta) = n( \operatorname{Re}(z_n)
\operatorname{Im}(z_n) ) + (n+1/2)
\arg(z_n)$ [recall (\ref{mainexpexpand})], there is the estimate
\[
e^{i f_n(r,\theta)} \widetilde{S}_n (r, \theta; s, \eta)
e^{- i f_n(s,
\eta)} = \frac{1}{2 \pi} \sqrt{\frac{\gamma_n}{n}}
\frac{ e^{-(r+s)/2} }{ (1 + \sqrt{\gamma_n/n}) e^{i (\theta
-\eta)} -1 } \bigl(1 + o(1) \bigr),
\]
in the ``bulk'' of $\mathcal{T}$,
as advertised in (\ref{Skappa}), with $\kappa= \sqrt{\frac{\gamma_n}{n}}$.
\end{rem*}

\textit{Step} 3 considers again $r, s \le{n}^{1/4}$, but keeps
$\varepsilon$ and $\eta$ away from the diagonal via $|\theta- \eta| >
\varepsilon$,
the latter defined in (\ref{epschoice}). With now
$ | z_n \bar{w}_n - 1| \ge\frac{1}{C} \varepsilon$ [see (\ref{tau})]
we have that
%
\begin{eqnarray}
\label{outerbound}
&& \bigl| e^{-n (\operatorname{Im}(z_n)^2 + \operatorname{Im}(w_n)^2)}  e^{-(n/2) (z_n
-\bar{w}_n)^2} e^{n z_n \bar{w}_n}
\mathfrak{e}_{n-2}( n z_n \bar{w}_n) \bigr|\nonumber
\\
&&\qquad \le {C} \frac{1}{\varepsilon} \exp \biggl( n \biggl(1 - \frac{1}{2}
|z_n|^2 - \frac{1}{2} |{w}_n|^2
+ \log|z_n\|w_n| \biggr) \biggr)
\\
&&\qquad \le C \frac{\gamma_n}{n \varepsilon} e^{ - (r+s)/8}.\nonumber
\end{eqnarray}
Lemma \ref{LKKMM} is responsible for the first inequality, producing
the same exponent as in step~2. Line 2 then reuses estimate
(\ref{mainexpbound})
from that step in conjunction with~(\ref{scalingdef}).

Next, while the considerations behind (\ref{prefactor}) still hold,
here the $\sin( \frac{\theta+ \eta}{2}) + \varepsilon$ is of little
use. Instead we are lead to the bound
%
\begin{eqnarray}\label{bigrbound}
\bigl| \widetilde{S}_n (r,\theta; s, \eta) \bigr| &\le& C \sqrt{
\frac{\gamma_n}{n} } \frac{1}{\varepsilon} \frac{
e^{-(r+s)/8} }{\sqrt{\sin(\theta) \sin(\eta)}}
\nonumber\\[-8pt]\\[-8pt]
\eqntext{\mbox{on } |\theta-\eta| > \varepsilon\mbox{ and } r, s \le n^{1/4}.}
\end{eqnarray}
First keeping $\theta$ and $\eta$ away from the origin, let
\[
\mathcal{O}_n= \bigl\{ \varepsilon< \theta, \eta< \pi/2, | \theta-
\eta| > \varepsilon r,s \le n^{1/4} \bigr\}
\]
for which we have that
\[
\int_{\mathcal{O}_n} |\widetilde{S}_n |^2 \le
C \frac{\gamma_n}{n
\varepsilon
^2} \biggl(\int_{\varepsilon}^{\pi/2}
\frac{ d \theta}{\sin
(\theta)} \biggr)^2 \le C \frac{\gamma_n^4}{n^{1/4}},
\]
having substituted (\ref{epschoice}). The same bound holds for the
integral over the set analogous to $\mathcal{O}_n$ but with $\theta,
\eta< \pi- \varepsilon$.

To finish, as in step~2 we control the integral over the region where
say $0 \le\theta< \varepsilon$ by altering our initial bound on the
$\phi$ function(s). Here though this is done in just the variable near
the singular point. To illustrate, the bound
\[
\bigl| \widetilde{S}_n (r,\theta; s, \eta) \bigr| \le C \sqrt{
\frac{\gamma_n}{n} } \frac{n^{1/4}}{\varepsilon} \frac{
e^{-(r+s)/8} }{\sqrt{\sin(\eta)}} \equiv C
\gamma_n \frac{ e^{-(r+s)/8}
}{\sqrt{\sin(\eta)}},
\]
is again valid throughout the region described in (\ref{bigrbound}),
but useful only when $\theta$ is small where it produces
\[
\int_{ \mathcal{O}_n \cap\{\theta< \varepsilon\} } | \widetilde{S}_n|^2 \le
C \gamma_n^2 \int_0^\varepsilon\bigl|
\log(\theta+ \varepsilon) \bigr| \,d \theta\le C \frac{\gamma_n^2}{n^{1/4}}.
\]
Once more, like considerations apply to $\theta$ near $\pi$ (and also
of course to the situation where $\theta$ and $\eta$ change roles).

\textit{Step} 4 dispenses of the case that either $r$ or $s$ is greater
than ${n}^{1/4}$. Once again
$|z_n \bar{w}_n -1 | \ge| |z_n \bar{w}_n| -1 | \ge C \varepsilon$ and
\[
\bigl| e^{-n (\operatorname{Im}(z_n)^2 + \operatorname{Im}(w_n)^2)} e^{-(1/2) (z_n -\bar
{w}_n)^2} e^{n z_n \bar{w}_n} \mathfrak{e}_{n-2}(
n z_n \bar{w}_n) \bigr| \le C \frac{\gamma_n^{1/2}}{n^{3/4} }
e^{ - (r+s)/8},
\]
exactly as in (\ref{outerbound}), now just employing the definition of
$\varepsilon$. The relevant bound on the kernel becomes
\[
\bigl| \widetilde{S}_n (r,\theta; s, \eta) \bigr| \le C n^{1/4} {
\gamma_n} \bigl(1+ |r| + |s|\bigr) e^{ - (r+s)/8}.
\]
Here we have again used the simplified bound $\phi(a) \le e^{-a^2/2}$,
as well as the even rougher estimate $|z_n - \bar{w}_n| \le(C + |r| +
|s|)$ in the prefactor. In any case it is enough. The square integral
of the above restricted to $\{ r \vee s > n^{1/4} \}$ is dominated by
$C e^{-n^{1/4}/C}$.

\textit{Step} 5 is to note that everything above applies to $\widetilde{DS}_n$
(or $\widetilde{IS}_n$) with one notable change. The appearance of
$(w_n -
z_n)$ [or $(\bar{w}_n - \bar{z}_n)$] in (\ref{prefactor}) rather than
$(\bar{w}_n - z_n)$, leads to the replacement of
the factor $ [ \sin( \frac{\theta+ \eta}{2}) + \varepsilon]$ with
$[\sin( \frac{\theta- \eta}{2}) + \varepsilon]$. The latter is
$O(\varepsilon)$ on any $ \mathcal{N}_{n, \theta}$, producing an
additional decay along the diagonal. This completes the proof.
\end{pf*}

We close this section with the following:

\begin{pf*}{Proof of Corollary \ref{compcor}}
One needs to show that,
for any nonnegative $f(r, \theta)$ supported on $\{r > t\}$,
%
\begin{equation}
\label{laplacefunct} \lim_{n \rightarrow\infty} \mathbb{E} \biggl[ \prod
_{z_k \in
\mathbb{C}_+} e^{ - f
(r_k', \theta_k')} \biggr] = \exp \biggl[- \int
_t^{\infty}\!\! \int_0^\pi
\bigl(1- e^{- f(r, \theta)} \bigr) \frac{1}{2 \pi} e^{-r} \,dr \,d \theta
\biggr],\hspace*{-30pt}
\end{equation}
recall the scaling $z_k \mapsto z_k' =(r_k', \theta_k')$ from the
statement. But, by the above, the square of the expectation on the left
is $\det(I - K_n (1- e^{-f} ))$. Since $|1- e^{-f}|$ is bounded by
$\chi$, all the estimates in the previous proof apply with the result
being the exponential of the trace of
$-K_n (1- e^{-f} )$. This is exactly the right-hand side~(\ref{laplacefunct}).
\end{pf*}

\section{Real points}\label{sec4}

We run through the calculation over even values of $n$, returning to
the modifications required for $n$ odd at the end.

\subsection{$n$ even}\label{sec4.1}

First the determinant of (\ref{realK}) at finite $n$ is reduced to that
of a~scalar operator. Throughout this section
any $\chi$ appearing on its own denotes $\chi= \chi_{\{ t < x <
\infty
\} }$.

%
%
\begin{lemma}
\label{TWmoves}
With $K_n$ defined in (\ref{realK}) we have that
%
\begin{equation}
\label{realdetscalar} \det(I - K_n \chi) = \det(I - T_n \chi)
\det(I - W_n).
\end{equation}
Here $T_n$ is the symmetric part of $S_n$ [recall (\ref{realS}),
%
\begin{equation}
T_n(x,y) = \frac{e^{-(1/2) (x-y)^2}}{\sqrt{2 \pi}} e^{-xy} \mathfrak
{e}_{n-2}(xy)]
\end{equation}
and $W_n$ is a finite rank operator defined in~(\ref{eq2}).
\end{lemma}

This step in particular mimics Tracy and Widom's treatment of GOE
[\citet{MR1657844}] quite closely. Next, introducing the scaling as in
%
\begin{equation}
\label{scalingT} \widetilde{T}_n(x,y) = T_n(\sqrt{n} + x,
\sqrt{n} + y ),
\end{equation}
the convergence of the first factor in (\ref{realdetscalar}), and more,
is dealt with by the following [the point being that $\det(I -
\widetilde{T}_n \chi) = \det( I - \chi\widetilde{T}_n \chi) $].

%
%
\begin{lemma}
\label{Tconvergence} For all $t > -\infty$, the $L^2$ operator $ \chi
\widetilde{T}_n \chi$ converges in trace norm to $T \chi$ with the
kernel for $T$
defined in (\ref{Tkernel}). Further, $ \chi\widetilde{T}_n \chi
\rightarrow\chi{T}_n \chi$ and $(I - \chi\widetilde{T}_n \chi
)^{-1} \rightarrow(I- \chi T \chi)^{-1} $ in
$L^1, L^2$ and $L^{\infty}$
operator norms.
\end{lemma}

The last step deals with the $W_n$ operator appearing in the second
factor of (\ref{realdetscalar}). The proof of Lemma \ref{TWmoves} will
show that $W_n$ is of the form
%
\begin{equation}
\label{eq2} W_n = \alpha_1 \otimes
\beta_1 + \alpha_2 \otimes\beta_2
\end{equation}
in which
\begin{eqnarray*}\label{alphabetas}
\alpha_1 &=& (I -T_n \chi)^{-1} \phi_n,\qquad \beta_1 = \chi\psi_n,
\\
\alpha_2 &=& \tfrac{1}{2} \bigl( (
\psi_n, I -\chi) (I - T_n \chi)^{-1}
\phi_n+ (I - T_n \chi)^{-1} T_n (I -
\chi) \bigr),
\\
\beta_2 &=& \delta_t -\delta_\infty,
\end{eqnarray*}
$\phi_n(x) = \kappa_n \int_0^x u^{n-2} e^{-(1/2) u^2}   \,du
$ and
$\psi_n(x) = \kappa_n' x^{n-1} e^{-(1/2) x^2}$ (with certain
constants $\kappa_n, \kappa_n'$).
The determinant of $I - W_n$ is then comprised explicitly of the
\mbox{$L^2$-}inner products $(\alpha_i, \beta_j)_{1\le i,j \le2}$, and what
we need is the following:

%
%
\begin{lemma}
\label{Wconvergence}
After scaling as in (\ref{scalingT}), the inner products $(\alpha_i,
\beta_j)_{1\le i,j \le2}$ converge to their formal limits.
\end{lemma}

The object identified as $\Gamma_t$ in the statement of Theorem \ref{RealThm} is just the expansion of ``$\det(I - W_{\infty})$.''

Before the proofs of Lemmas \ref{TWmoves}, \ref{Tconvergence} and
\ref{Wconvergence}, we verify, as indicated in the \hyperref[sec1]{Introduction}, that the
kernel $T$
does not have a ``Christoffel--Darboux'' structure.
G\'erard Letac showed us this short argument.
The question is whether there exist functions $F$ and $G$ (which can
assumed to be $C^2$) for which
\[
\int_0^{\infty} e^{-(x+u)^2} e^{-(y+u)^2}
\,du = \frac{ F(x) G(y) - F(y)
G(x)}{x-y}.
\]
The answer is no. Since the left-hand side is of the form $e^{-x^2}
e^{-y^2} H(x+y)$, it is enough to prove that
\[
H(x+y) = \frac{ F(x) G(y) - F(y) G(x)}{x-y}
\]
is impossible except for $F$ and $G$ proportional, and so $H =0$.
Making the change of variables $t=x-y$ and $s=x+y$ we find that
\[
\frac{\partial}{\partial t}{\frac{1}{t}} \bigl( F(s+t) G(s-t) - F(s-t) G(s+t)
\bigr) = 0.
\]
This implies that
\[
t \mapsto F(s+t) G(s-t) - t F(s+t) G'(s-t) + t
F'(s+t) G(s-t)
\]
is an even function. Differentiating this function with respect to $t$
and setting $t=0$ yields $F'(s) G(s) = F(s) G'(s)$ which was the claim.

\begin{pf*}{Proof of Lemma \ref{TWmoves}}
We start with the matrix kernel
%
\begin{equation}
\label{eq7} K_n = \lleft[\matrix{ S_n & \delta
S_n^{\mathsf{T}}
\vspace*{3pt}\cr
- \epsilon S_n + \epsilon&
S_n^{\mathsf{T}} } \rright].
\end{equation}
This is exactly analogous to the kernel for GOE given by \citet
{MR1657844}, and we follow the strategy laid out in Section~II of that paper.
First, since $\delta\epsilon= -I$,
\[
\chi K_n \chi= \lleft[\matrix{ \chi\delta& 0
\vspace*{3pt}\cr
0 & \chi } \rright]
\lleft[\matrix{ - \epsilon S_n \chi& S_n^{\mathsf{T}}
\chi
\vspace*{3pt}\cr
(-\epsilon S_n + \epsilon) \chi& S_n^{\mathsf{T}}
\chi } \rright].
\]
Using the famous $\det(\mathbf I - \mathbf{AB}) = \det(\mathbf I
- \mathbf{BA})$ trick, we find the
Fredholm determinant of $\chi K_n \chi$ equals that of
\[
\lleft[\matrix{ -\epsilon S_n \chi& S_n^{\mathsf{T}}
\chi
\vspace*{3pt}\cr
(-\epsilon S_n + \epsilon) \chi& S_n^{\mathsf{T}}
\chi } \rright] \lleft[\matrix{ \chi\delta& 0
\vspace*{3pt}\cr
0 & \chi } \rright] =
\lleft[\matrix{ -\epsilon S_n \chi\delta& S_n^{\mathsf{T}}
\chi
\vspace*{3pt}\cr
-\epsilon S_n \chi\delta+ \epsilon\chi\delta&
S_n^{\mathsf{T}} \chi } \rright].
\]
The determinant is further unaffected if we subtract the first row from the
second and then add the second column to the first, resulting in
\[
\lleft[\matrix{ S_n^{\mathsf{T}} \chi- \epsilon
S_n \chi\delta& S_n^{\mathsf{T}} \chi
\vspace*{3pt}\cr
\epsilon\chi
\delta& 0 } \rright].
\]
The best way to understand this is to note this pair of moves is
affected by $K_n \mapsto P K_n P^{-1}$ with
$P = \bigl(
{\fontsize{8.36}{8}\selectfont{\matrix{1 &\!\! - 1\cr 0 &\!\! 1}}} \bigr)$.
Thus
\begin{eqnarray*}
\mathbb{P}_{n, \mathbb{R}}(t)^2 &=& \det\lleft[\matrix{ I -
S_n^{\mathsf{T}} \chi+ \epsilon S_n \chi\delta&
-S_n^{\mathsf{T}
} \chi
\vspace*{3pt}\cr
- \epsilon\chi\delta& I } \rright]
\\
&=& \det\lleft[\matrix{ I - S_n^{\mathsf{T}} \chi+ \epsilon
S_n \chi\delta- S_n^{\mathsf{T}} \chi\epsilon\chi
\delta& 0
\vspace*{3pt}\cr
0 & I } \rright],
\end{eqnarray*}
which follows by ``row reducing'' the matrix. And since one may check that
\mbox{$\epsilon S_n = S_n^{\mathsf{T}} \epsilon$,} we find that for even
$n$,
%
\begin{eqnarray}
\label{eqscalardets1} \mathbb{P}_{n, \mathbb{R}}(t)^2 
& =& \det \bigl( I - S_n^{\mathsf{T}} \chi+
S_n^{\mathsf{T}}(1 - \chi) \epsilon\chi\delta \bigr).
\end{eqnarray}
The above manipulations have been carried out completely formally, with
no attention as to in which space(s) the operators/determinants reside.
The needed technical details may be taken (yet again) verbatim from
\citet{MR1657844}; see Section~VIII.

Now for even $n$ recall the kernel
\begin{eqnarray*}
S_n^{\mathsf{T}}(x,y) & =& \frac{1}{\sqrt{2 \pi}} e^{-(1/2)(x-y)^2}
e^{-xy} \mathfrak{e}_{n-2}(xy)
\\
&&{} + \frac{1}{\sqrt{2 \pi}}
\frac{ y^{n-1}
e^{-y^2/2}}{(n-2)!} \int_0^{x}
u^{n-2} e^{-(1/2) u^2} \,du
\\
& =& T_n(x,y) + U_n(x,y),
\end{eqnarray*}
where we have previously defined $T_n$ as the symmetric part of $S_n$,
and consider now $U_n$ the remainder. For later it will be useful to
express $U_n$ (as an operator) as
%
\begin{equation}
\label{eqpsiphi}\quad  U_n = \phi_n \otimes\psi_n =
\biggl( \kappa_n \int_0^x
u^{n-2} e^{-(1/2) u^2} \,du \biggr) \otimes \bigl( \kappa_n'
y^{n-1} e^{-(1/2) y^2} \bigr),
\end{equation}
where $\kappa_n = \sqrt{ \frac{n^{1/2}}{\sqrt{ 2 \pi} (n-2)! }}$ and
$\kappa_n' = \sqrt{ \frac{n^{-1/2}}{\sqrt{ 2 \pi} (n-2)! }}$. Keep in
mind of course that $\phi_n= \epsilon\phi_n'$, $\phi_n'(x) = \kappa_n
x^{n-2} e^{-(1/2) x^2}$.

Next, introduce the resolvent
%
\begin{equation}
\label{TtoR} (I - T_n \chi)^{-1} = I + R_n\quad\mbox{or}\quad R_n = T_n \chi( I - \chi T_n
\chi)^{-1}
\end{equation}
with kernel $R_n(\cdot, \cdot)$. The determinant (\ref{eqscalardets1})
factors as in
%
\begin{eqnarray}
\label{eqscalardets2} \mathbb{P}_{n, \mathbb{R}}(t)^2 &= & \det(I -
T_n\chi)\det \bigl(I - \bigl((I - T_n\chi)^{-1}
\phi_n \bigr) \otimes( \chi\psi _n)
\nonumber\\[-8pt]\\[-8pt]
&&\hspace*{81pt}{} + (I - T_n
\chi)^{-1} S_n^{\mathsf{T}}(1 - \chi)\epsilon\chi\delta
\bigr),\nonumber
\end{eqnarray}
having used $A ( B \otimes C) D = (AB) \otimes(D^{\mathsf{T}} C)$.
The second term (in which $\epsilon\chi\delta$ appears) is simplified
by considering the commutator
\[
[\chi, \delta ] = - ( \delta_t \otimes\delta_t - \delta
_{\infty} \otimes\delta_{\infty}),
\]
where $\delta_a $ is the dirac delta.
Since again $\epsilon\delta= -I$,
\[
(1-\chi) \epsilon [ \chi, \delta ] = (1 -\chi) \epsilon \chi\delta= - (1 -\chi)
( \epsilon_t \otimes\delta_t - \epsilon _{\infty}
\otimes\delta_{\infty})
\]
with now
\[
\epsilon_t(x) = \tfrac{1}{2} \operatorname{sgn}(t - x) \qquad
\bigl(\mbox{so } \epsilon _{\infty}(x) \equiv\tfrac{1}{2} \bigr).
\]
Thus
\[
(I + R_n) S_n^{\mathsf{T}} (I - \chi) \epsilon\chi
\delta= - \tfrac
{1}{2} (I + R_n) S_n^{\mathsf{T}}
(1 - \chi) \otimes( \delta_t - \delta_{\infty} ).
\]
Expanding out the $S_n$ and noting
$ (A \otimes B) ( C \otimes D) = (B,C) (A \otimes D)$, the second
factor in (\ref{eqscalardets2}) is the determinant of
the identity minus the finite rank operator
\begin{eqnarray*}
W_n &= & \bigl( (I - T_n \chi)^{-1}
\phi_n \bigr) \otimes( \chi\psi_n)
\\
&&{} + \tfrac{1}{2} \bigl( (\psi_n, 1 -\chi) (I - T_n
\chi)^{-1} \phi _n + (I - T_n
\chi)^{-1} T_n (1 - \chi) \bigr) \otimes(
\delta_t - \delta _\infty),
\end{eqnarray*}
to which we apply the well-known fact
\[
\det(I - W_n) = \det \bigl(\delta_{i,j} - (
\alpha_i, \beta_j) \bigr)_{1\le i,j\le2}
\]
with, as announced above,
%
\begin{eqnarray}\label{alphasbetas}
\alpha_1 &=& (I - T_n \chi)^{-1}
\phi_n,\qquad  \beta_1 = \chi \psi_n,\nonumber
\\
\alpha_2 &=& \tfrac{1}{2} \bigl( (
\psi_n, 1 -\chi) (I - T_n \chi)^{-1}
\phi_n+ (I - T_n \chi)^{-1} T_n (1-
\chi) \bigr),
\\
\beta_2 &=& \delta_t - \delta_\infty.\nonumber
\end{eqnarray}
Here $(\alpha_i, \beta_j)$ are regular $L^2$-inner products.
(We have reused $\delta$ many times---here it is the standard
Kronecker delta.)
In particular, all components comprising the original $\det(I - K_n
\chi
)$ are well defined.

We conclude with a few simplifications. First,
%
\begin{eqnarray}
\label{row1} (\alpha_1, \beta_1) & =& \bigl( \chi
\phi_n, (I - \chi T_n \chi)^{-1} \psi
_n \bigr),
\nonumber\\[-8pt]\\[-8pt]
(\alpha_1, \beta_2) & =& \bigl( (I - T_n
\chi)^{-1} \phi_n \bigr) (t) - \phi_n(\infty).\nonumber
\end{eqnarray}
In the second line $(I - T_n \chi)^{-1} \phi_n = \phi_n + T_n \chi
(I -
\chi T_n \chi)^{-1} \phi_n$ is used.
Next, since $ ( T_n + T_n R_n^{\mathsf{T}} ) \chi= R_n$, it follows that
\begin{eqnarray*}
\bigl( (I+R_n) T_n (1- \chi), \chi\psi_n
\bigr) &=& (1 - \chi, R_n \psi_n),
\\
\bigl((I+R_n) T_n (1 - \chi), \delta_t \bigr) &=&
\bigl( I - \chi, R_n(\cdot, t) \bigr)
\end{eqnarray*}
and, with $c_n = (\psi_n, 1 -\chi) $,
%
\begin{eqnarray}
\label{row2} (\alpha_2, \beta_1) & =&
\frac{1}{2} \biggl( c_n (\alpha_1, \beta
_1) - c_n + \int_{-\infty}^t
(I+R_n) \psi_n(x) \,dx \biggr),
\nonumber\\[-8pt]\\[-8pt]
(\alpha_2, \beta_2) & =& \frac{1}{2} \biggl(
c_n ( \alpha_1, \beta _2) + \int
_{-\infty}^t R_n(x,t) \,dx \biggr).\nonumber
\end{eqnarray}
Combining (\ref{row1}) and (\ref{row2}), $\det(I - W_n) $ equals
\[
\bigl(1 - (\alpha_1, \beta_1) \bigr) \biggl( 1 -
\frac{1}{2} \int_{-\infty}^t
R_n(x,t) \,dx \biggr) - \frac{1}{2} (\alpha_1,
\beta_2) \int_{-\infty}^t
(I+R_n) \psi_n(x) \,dx;
\]
note the $c_n$ factor has dropped out. This last expression has
precisely the same structure as equation (41) in \citet{MR1657844}.
\end{pf*}

\begin{pf*}{Proof of Lemma \ref{Tconvergence}}
We employ the trace
convergence criteria of Theorem~2.20 of \citet{MR2154153}, showing that
%
\begin{equation}
\label{diagTconv} \int_{t}^{\infty}
\widetilde{T}_n(x,x) \,dx \rightarrow\int_t^{\infty}
T(x,x) \,dx
\end{equation}
and
%
\begin{equation}
\label{weakTconv} \qquad\int_t^{\infty}\!\! \int
_t^{\infty} f(x) \widetilde{T}_n(x,y) g(y)
\,dx \,dy \rightarrow\int_t^{\infty}\!\! \int
_t^{\infty} f(x) T(x,y) g(y) \,dx \,dy
\end{equation}
for all $f,g \in L^2([t,\infty))$. Both follow from the pointwise
convergence of $\widetilde{T}_n$ to $T$, and an easy domination.

Though $\widetilde{T}_n \rightarrow T$ pointwise already demonstrated
in \citet{borodin-2008}, it is useful here to establish some local
uniformity. Start with the expression
\[
\widetilde{T}_n(x,y) = \frac{e^{-(1/2) (x-y)^2}}{\sqrt{2 \pi}} e^{-n \eta
_n}
\mathfrak{e}_{n-2}(n \eta_n ), \qquad\eta_n =
\eta_n(x,y) = 1 + \frac
{x+y}{\sqrt{n}} + \frac{xy}{n}.
\]
Repeating the estimate
from Lemma \ref{wimp} (as used in step~1 of Lemma \ref{NormConvergence}) yields
\[
\widetilde{T}_n(x,y) = \frac{e^{-(1/2) (x-y)^2}}{ 2 \sqrt{
\pi}} \frac
{\mu(\eta_n) \eta_n}{\eta_n -1}
\operatorname{erfc} \bigl( \sqrt{n} \mu ( \eta_n) \bigr) \biggl( 1+ O
\biggl( \frac{1}{\sqrt{n}} \biggr) \biggr),
\]
uniformly, granted that $\eta_n \ge0$ which holds say for $x$ and $y$
bounded and $n$ large enough.
Since $\sqrt{n} \mu(\eta_n) \rightarrow\frac{x+y}{\sqrt{2}}$ for
uniformly for ${x}$ and $y$ on compacts, it follows $ \widetilde{T}_n(x,y)
\rightarrow\frac{e^{-(1/2) (x-y)^2}}{ 2 \sqrt{2 \pi}}
\operatorname{erfc}
( \frac{x+y}{\sqrt{2}}  )$ in the same fashion. A change of
variables shows this object is equivalent to $T$.
By the smoothness of the functions involved, we also have a constant
$C$ so that $\widetilde{T}_n(x,y) \le C T(x,y)$ for all $x$ and $y$ bounded.

For the rest of domination, on diagonal $\eta_n = (1 + x/\sqrt{n})^2$
is always nonnegative, and we can continue as above. In particular if
$x \ge1$,
\[
\frac{\mu(\eta_n) \eta_n}{\eta_n -1} \operatorname{erfc} \bigl( \sqrt {n} \mu( \eta
_n) \bigr) \le \frac{\eta_n}{\sqrt{n}(\eta_n-1)} e^{-n \mu^2(\eta_n)} \le x
e^{- x^2},
\]
since $\mu^2((1+a)^2) \ge a^2$. This is enough to conclude that (\ref
{diagTconv}) holds. Off diagonal, let $x+y \ge1$, go back to the
definition of $\widetilde{T}_n(x,y)$, and note quite simply that $| e^{-n
\eta
_n} \mathfrak{e}_{n-2} (n \eta_n) | \le e^{ - n \eta_n + n | \eta_n|
}$. Hence, when $xy>0$ as well we have that $\widetilde{T}_n(x,y) \le
e^{-(1/2) (x-y)^2}$ which controls that range of the integral in
(\ref{weakTconv}): $\int e^{-(1/2) (x-y)^2} f(y) \,dy \in L^2$
for $f \in L^2$. On the other hand, if, for instance, $x>0$
and $y <0$ (requiring $t <0$), the same observation gives $\widetilde
{T}_n(x,y) \le e^{-(1/2) x^2} e^{4|t| x}$ which suffices for the
remaining variable range in~(\ref{weakTconv}).

The trace norm convergence certainly implies the $L^2$ operator norm
convergence of $\chi\widetilde{T}_n \chi$. More directly though, any
symmetric kernel operator of the form $\chi M \chi$ has $L^p \mapsto
L^p$ norm, for $p=1,2, \infty$ bounded as in
\[
\| \chi M \chi\| \le\sup_{y> t} \int_t^{\infty}
\bigl|M(x,y)\bigr| \,dx;
\]
the $L^2 \mapsto L^2$ bound, less familiar than the $L^2 \otimes L^2$
kernel norm bound, is due to Holmgren [\citet{MR1892228}, Section~16.1]. The estimates above will then imply that $\chi\widetilde{T}_n
\chi
\rightarrow\chi T \chi$ in the $L^{1}$ and $L^{\infty}$ operator norms
too. As for the resolvents, it is enough to check that
\begin{eqnarray*}
\sup_{y> t} \int_t^{\infty} T(x,y)
\,dx & \le&\sup_{y> t} \frac
{1}{\sqrt { \pi}} \int
_0^{\infty} e^{-(u+y)^2} \,du =
\frac{1}{\sqrt{ \pi}} \int_{
t}^{\infty}
e^{- u^2 } \,du < 1
\end{eqnarray*}
for $t > - \infty$.
\end{pf*}

\begin{pf*}{Proof of Lemma \ref{Wconvergence}}
The scaling is to shift $t$ by $ \sqrt{n}$ in all appearances of~$\chi$
(and so $R_n$). This shift then filters into $\phi_n, \psi_n$ and so on
by changing variables in each $(\alpha_i, \beta_j)$. Again all scaled
functions/operators are decorated with tildes.

With now $g(x) = \frac{1}{\sqrt{2\pi}} e^{-x^2/2}$, $G(x) = \int_{-\infty}^x g(s) \,ds$ one readily finds: pointwise as $n \rightarrow
\infty$,
%
\begin{equation}
\label{scalarconv} \tilde{\psi}_n(x) \rightarrow g(x), \qquad\tilde{
\phi}_n(x) \rightarrow G(x).
\end{equation}
The first convergence also takes place in $L^1 \cap L^2$, while the
latter can be considered to hold in $\mathbb{R}+ L^2(s,\infty)$ for
any $s >
-\infty$ upon writing
$ \phi_n(x) = \phi_n(\infty) - \int_x^{\infty} \phi_n'$.

Allied considerations will also show that
%
\begin{equation}
\label{kernelconv} \widetilde{T}_n(x,t) \rightarrow T(x,t), \qquad\int
_{-\infty
}^{\infty} \widetilde {T}_n(x,y) \,dy
\rightarrow\int_{-\infty}^{\infty} {T}(x,y) \,dy,
\end{equation}
pointwise and in $(L^1 \cap L^2)(t,\infty)$.

Starting with $(\alpha_1, \beta_1)$ we find that
\begin{eqnarray*}
\label{firstinnerprod} (\tilde{\alpha}_1, \tilde{ \beta}_1) &
=& \bigl( \tilde{\phi}_n \chi, ( I - \chi\widetilde{T}_n
\chi)^{-1} \tilde{\psi}_n \bigr)
\\
& =& \bigl( \tilde{\phi}_n(\infty) \chi, \tilde{\psi}_n
\bigr) - \bigl( \bigl( \tilde{\phi}_n( \infty) - \tilde{
\phi}_n \bigr) \chi, ( I - \chi\widetilde{T}_n
\chi)^{-1} \tilde{\psi}_n \bigr).
\end{eqnarray*}
The convergence of $ \tilde{\phi}_n(\infty) \rightarrow1$ can
be viewed
as holding uniformly, and, by the second item in (\ref{scalarconv}), $
\tilde{\phi}_n(\infty) - \tilde{\phi}_n(\cdot)$
converges in $L^2(t,
\infty
)$. Further,
$( I - \chi\widetilde{T}_n \chi)^{-1} \tilde{\psi}_n(\cdot)
\rightarrow(I -
\chi T \chi)^{-1} g(\cdot)$ in $L^1 \cap L^2$ by the first item in
(\ref{scalarconv}) and Lemma \ref{Tconvergence}.

The next three items are treated similarly (to each other). Take\break  $\int_{-\infty}^t R_n(x,t) \,dx$, and rewrite the scaled version as
%
\begin{eqnarray}\label{RR}
&& \int_{-\infty}^t
\widetilde{R}_n(x,t) \,dx
\nonumber\\[-8pt]\\[-8pt]
&&\qquad = \int_{-\infty}^t
\widetilde{T}_n \chi(I - \chi\widetilde{T}_n
\chi)^{-1} \widetilde{T}_n(x, t) \,dx + \int
_{-\infty}^t \widetilde {T}_n(x,t) \,dx.\nonumber
\end{eqnarray}
The first term is the $L^2(t,\infty)$ inner product of the functions
%
\begin{equation}
\label{innerTS} \int_{-\infty}^t
\widetilde{T}_n(x, \cdot) \,dx \quad\mbox{and}\quad(I - \chi
\widetilde{T}_n \chi)^{-1} \widetilde{T}_n(
\cdot, t),
\end{equation}
each of which converges in $L^2(t,\infty)$ by (\ref{kernelconv}). The
second term converges to $\int_{-\infty}^t {T}(x,t) \,dx$ also by (\ref
{kernelconv}).
[As in \citet{MR1657844}, it is most convenient to see this by writing
$\int_{-\infty}^t R = (\int_{-\infty}^{\infty} - \int_{t}^{\infty})
R_n$ before applying the identity inherent in (\ref{RR}).]

The term $\int_{-\infty}^t (I - T_n \chi)^{-1} \psi  \,dx$ is easier.
Now we have that
\[
\int_{-\infty}^t (I + \widetilde{T}_n
\chi)^{-1} \tilde{\psi }_n \,dx = \int_{-\infty
}^t
\widetilde{T}_n \chi(I - \chi \widetilde{T}_n
\chi)^{-1} \tilde{\psi}_n(x) \,dx + \int_{-\infty}^t
\tilde{\psi}_n(x) \,dx.
\]
The only real change is the replacement of $ (I - \chi\widetilde{T}_n
\chi
)^{-1} \widetilde{T}_n(\cdot, t)$, appearing in~(\ref{RR}), with $
(I -\chi\widetilde{T}_n \chi)^{-1} \tilde{\psi}_n(\cdot)$.
We already have noted that this tends to its formal limit in~$L^2$.
Finally, $\int_{-\infty}^t \tilde{\psi}_n(x) \,dx$ is the same as
$\tilde{\phi}_n(t)$ up to trivial factors and also converges to $G(t)$.

Returning to $(\alpha_1, \beta_2)$, we only need deal with
\[
\widetilde{T}_n \chi(I - \chi\widetilde{T}_n
\chi)^{-1} \tilde{\phi}_n(t) = \int_t^{\infty}
\widetilde{T}_n(t,x) (I - \chi \widetilde{T}_n \chi
)^{-1} \tilde{\phi}_n(x) \,dx.
\]
Again, one can decompose $\tilde{\phi}_n(x) = \tilde{\phi}_n(\infty) - (
\tilde{\phi}_n(\infty) - \tilde{\phi}_n(x))$ and alternatively use the
$L^1(t,\infty)$ or $L^2(t, \infty)$ convergence
of $\widetilde{T}_n(t, \cdot)$ coupled with the $L^{2}+ L^{\infty}$
convergence of $ (I - \chi\widetilde{T}_n \chi)^{-1} \tilde{\phi}_n(x) $.
\end{pf*}

\subsection{$n$ odd}\label{sec4.2}

By Proposition~\ref{prop1}, when $n$ is odd
\[
\mathbb{P}_{\mathbb{R}, n}(t)^2 = \det(I - K_n \chi),
\]
where
\[
K_n = \lleft[\matrix{ S_n & \delta
S_n^{\mathsf{T}}
\vspace*{3pt}\cr
- \epsilon S_{n-1} + \epsilon+ (
\phi_n \otimes\varphi_n - \varphi_n \otimes
\phi_n ) & S_n^{\mathsf{T}} } \rright],
\]
where $\phi_n$ is as in (\ref{eqpsiphi}) and $\varphi= \epsilon
\psi
_n$ with also $\psi$ as in (\ref{eqpsiphi}).

Performing the same maneuvers as for the even $n$ kernel, we find that
$\mathbb{P}_{\mathbb{R}, n}(t)^2$ ($n$ odd) equals the determinant of
\begin{eqnarray*}
&& I - S_n^{\mathsf{T}} \chi - \tfrac{1}{2}
S_n^{\mathsf{T}} (1 - \chi) \otimes(\delta_t - \delta
_{\infty})
\\
&&\qquad{} + S_n^{\mathsf{T}} \chi \bigl(S_n^{\mathsf{T}} -
S_{n-1}^{\mathsf{T}} \bigr)\epsilon\chi\delta+ S_n^{\mathsf{T}}
\chi(\phi_n \otimes\varphi _n - \varphi_n
\otimes\phi_n ) \chi\delta.
\end{eqnarray*}
Again we can factor out the $( I - T_n \chi)$, and are left with two
extra components as compared with (the second factor on the right-hand
side of) equation (\ref{eqscalardets2}).

To see that the first extra component, featuring $S_n^{\mathsf{T}} -
S_{n-1}^{\mathsf{T}}$, gives no contribution~to the determinant in the
limit note the following. From before
$ \chi(\widetilde{T}_n - \widetilde{ T}_{n-1} )\chi\rightarrow0$,
in trace as well
as the $L^1 $ and $L^{\infty}$ operator norms, so this piece may be
taken out as a perturbation to the (previously factored) $( I - T_n
\chi
)$. Also, we have that
$ \tilde{\phi}_n - \tilde{\phi}_{n-1} \rightarrow0$, and $
\tilde{\psi}_n
- \tilde{\psi}_{n-1} \rightarrow0 $
[pointwise and in $\mathbb{R}+ L^2(s,\infty)$ or $L^2$, resp.]. The
latter two differences will appear as factors in some appropriate
``$\alpha$'s'' and ``$\beta$'s'' in the limiting finite rank
determinant, so it is enough
that all $L^2$-inner products in which they figure are zero.

Similarly, writing
\[
\phi_n \otimes\varphi_n - \varphi_n \otimes
\phi_n = (\phi_n - \varphi _n) \otimes
\varphi_n - \varphi_n \otimes( \phi_n -
\varphi_n ),
\]
the vanishing of
$\tilde{\phi}_n - \tilde{\varphi}_n$ pointwise and in
$\mathbb{R}+ L^2(s,\infty
)$ is
sufficient to conclude that any limiting inner product in which these
terms enter in will also be zero.
This completes the verification.

\section{Numerics and open questions}\label{sec5}

Not having a closed form for the limiting distribution of the largest
real point prompted us to carry out some \mbox{straightforward} simulations of
the matrix ensembles, resulting in
a few notable observations surrounding this object, as well as the
finite $n$ behavior of both the largest real and (in absolute value)
complex points. (From now on we write just ``largest complex point or
eigenvalue''---that we mean in absolute value should be understood.)

%
%
\begin{figure}

\includegraphics{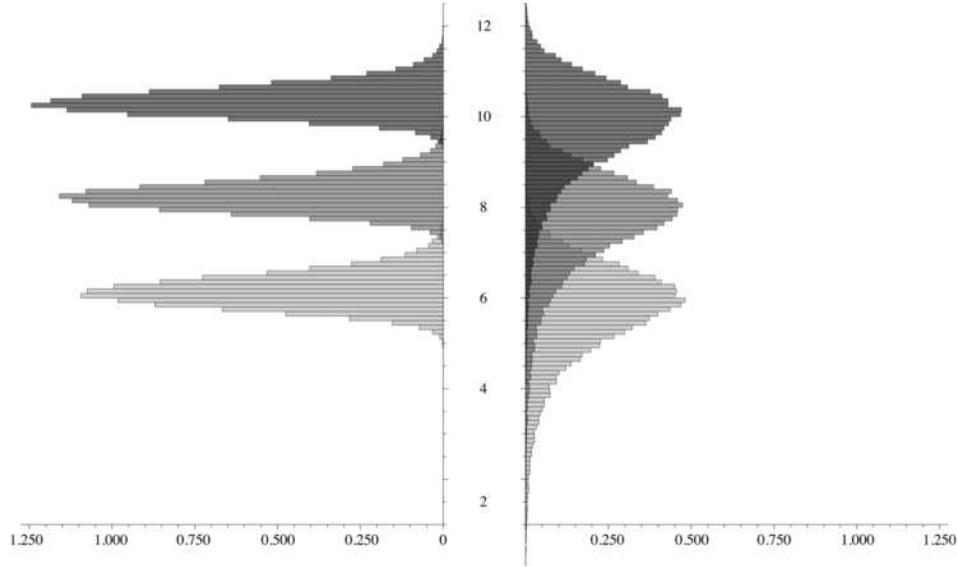}

\caption{Normalized histograms for the modulus of the largest
complex eigenvalue (left) and the largest real eigenvalue of 40,000
random $n \times n$ matrices for, from lightest to darkest, each
of $n=36, 64$ and 100.}\label{fig1}
\end{figure}

Figure~\ref{fig1} compares the histograms for the largest real and
complex points at $n = 36, 64$ and $100$.
Noticeable right away is the heavy left tail in the real point
distribution. One might have expected
that the tail going into the bulk of the spectrum would be lighter than
that held down simply by the Gaussian weight.
On the other hand, recall that there are only $O(\sqrt{n})$ real
eigenvalues for $n \uparrow\infty$ [\citet{MR1231689}], and so rather
weak level repulsion along the real line. As the right tail of this law
can be seen to have Gaussian decay, a reasonable
conjecture is that the limiting left tail is exponential to leading order.

\begin{figure}[b]

\includegraphics{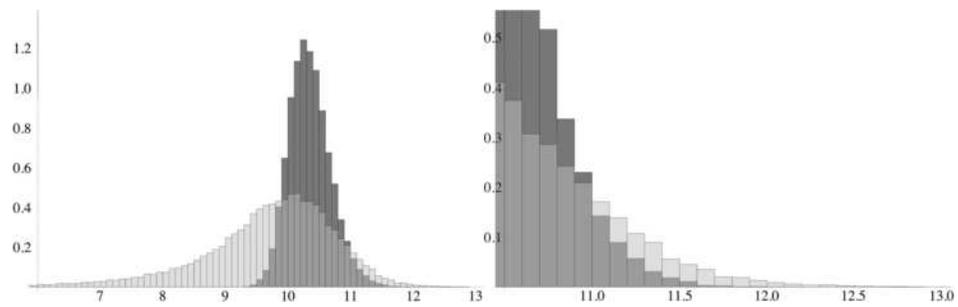}

\caption{Histograms of the largest complex (dark) and real (light)
eigenvalues of 40,000 random $100 \times100$ with an enlarged
picture of the right hand tails.}\label{fig4}
\end{figure}

A closer look at Figure~\ref{fig1} also shows that the empirical
distribution of the largest complex point appears far more symmetric
than its
limiting Gumbel shape would suggest. Figure~\ref{fig4} focuses in on
the $n=100$ case and highlights that at least for moderate $n$
the real point distribution is heavier tailed to the right as well.
This gets right into some basic questions on the speed of convergence for
these laws.

%
%

Experts of RMT anticipate fast convergence of finite $n$
statistics to their limiting distribution. Here though the Gumbel
shape of the spectral radius will not ``kick-in'' until $n$ is
considerably large. What dictates the phenomena is that the
real point is centered about $\sqrt{n}$ while the complex point is centered
about $\sqrt{n} + \sqrt{\gamma_n}$ with $\gamma_n = \log (
\frac{n}{2 \pi(\log n)^2}  )$. Even as written, this is a purely
asymptotic statement, $\gamma_n$ is not even positive until $n \approx
165$, and in any case one sees that $n$ has to be much larger still
until the competing real/complex distributions separate.

%
%
\begin{figure}

\includegraphics{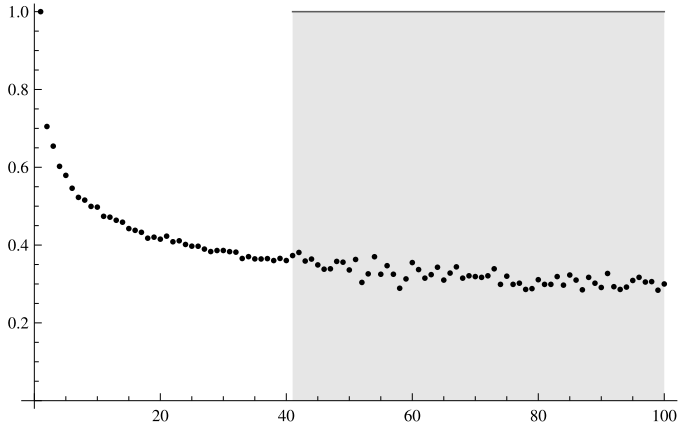}

\caption{The proportion of samples for which the largest eigenvalue
was real as a function of $n$. The unshaded region was sampled
10,000 times for each $n$; the shaded region 1000 times for each $n$.}\label{fig2}
\end{figure}

A first question might then be what is the chance, at finite $n$, that the
spectral radius comes from a real point? Figure~\ref{fig2} graphs this
empirical probability, indicating this
is a slowly decaying function, still just under 0.4 for $n=100$.
Quantifying this analytically
appears challenging. To start, one would need sharp control of the
mixed (real/complex) gap probability, given by the Fredholm
determinant/Pfaffian of a $4 \times4$ matrix kernel operator.

\begin{figure}[t]

\includegraphics{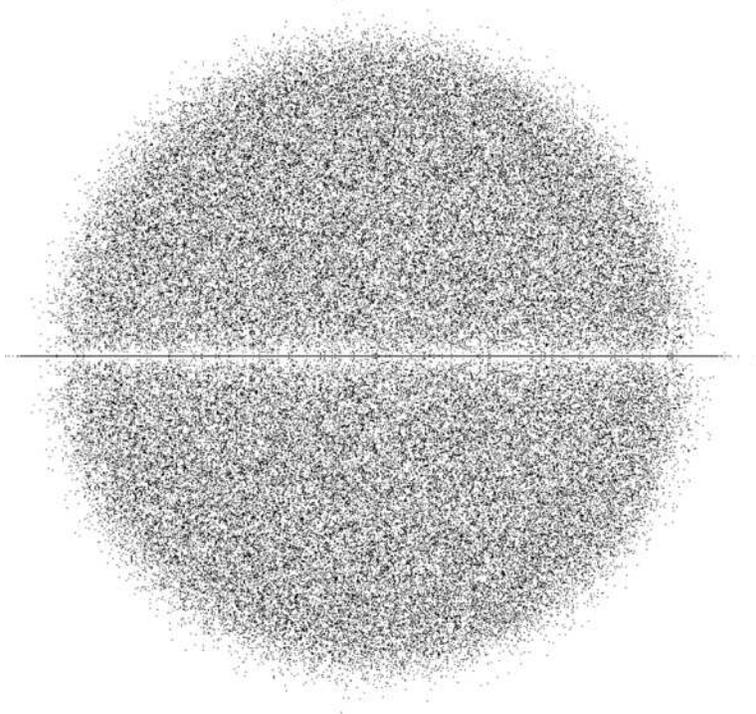}

\caption{A plot of the eigenvalues of 1000 random $100 \times100$
matrices.}\label{fig5}
\end{figure}

Together with Figure~\ref{fig4}, the phenomena appears to be as
follows. At moderate~$n$ (e.g., $n=100$), the largest eigenvalue is most
likely complex.
However, in the situation where the largest
eigenvalue \textit{is} real, it is more likely to be larger
than if it were complex. With hindsight one can see this in the most
famous picture
attached to the real Ginibre ensemble which we repeat an instance of
here in Figure~\ref{fig5}.
The striking feature is the so-called ``Saturn effect,'' based on which
alone a person might
be forgiven for having conjectured that the largest eigenvalue would be
real, with probability one,
as $n \uparrow\infty$. Rather, the Saturn effect is a phenomenon
which appears from plotting the eigenvalues of many matrices
simultaneously. Eventually, the complex points overwhelm the
$O(\sqrt{n})$ on the real line.

%
%

%
%
\begin{figure}[b]

\includegraphics{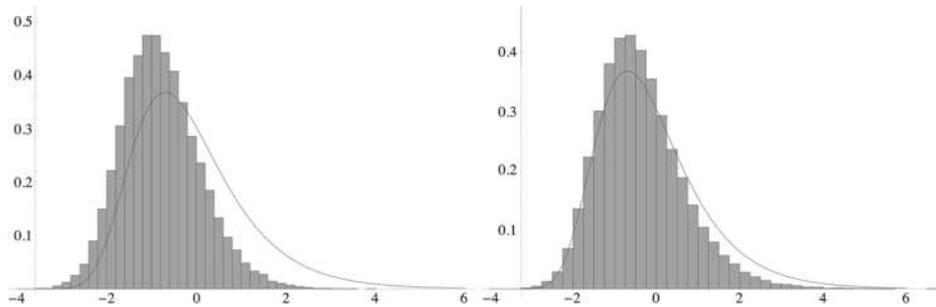}
\caption{A scaled and shifted histogram of the largest complex
eigenvalue (left) and the spectral radius (right) of
40,000 $100 \times100$
matrices as compared to the Gumbel density $\frac{1}{2}
e^{-t-(e^{-t}/2)}$.}\label{fig3}
\end{figure}

\begin{figure}
\includegraphics{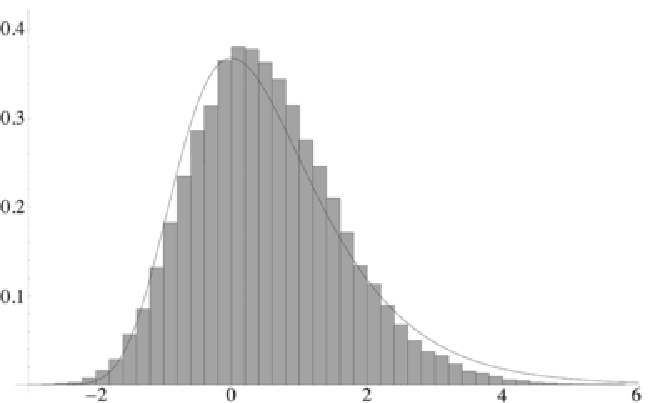}
\caption{A histogram of the spectral radius of 40,000 complex random
$100 \times100$ matrices as compared to the Gumbel density
$e^{-t-e^{-t}}$.}\label{fig6}
\end{figure}

In summary, one cannot expect the Gumbel law to be a good approximation
for the spectral radius at small $n$.
And this is due to more than just the mixture of the separate largest
real and complex point laws. Figures~\ref{fig1} and~\ref{fig4} already
show that the largest complex point distribution itself is not well
approximated by its limiting Gumbel. Figure~\ref{fig3}
makes this more transparent. The issue with $\sqrt{\gamma_n}$ not being
sensible for smaller $n$ is circumvented
by appealing to equation (\ref{scalingdef}), $ \lim_{n \rightarrow
\infty} \frac{\sqrt{n}}{\sqrt{2
\pi} \gamma_n} e^{-\gamma_n/2} = 1$, which more or less defines
$\gamma_n$. For\vspace*{-1pt} numerical comparisons then we take
$\gamma_n$ to be the solution $\gamma$ of $ \gamma^2 e^{\gamma} =
\frac
{n}{2 \pi} $.

More confounding in Figure~\ref{fig3} is that a scaled spectral radius
appears better approximated by the
Gumbel distribution than does the largest complex eigenvalue. This
though must be purely superficial.
It again comes back to the fact that when the largest eigenvalue is
real it tends to be
larger than were it complex, thickening the right tail
of this histogram to look more Gumbel. As emphasized many times,
however, this phenomenon
vanishes in the large $n$ limit.

In light of all this, a fair question that remains is
how to engineer a decent fluctuation approximation for the spectral
radius at finite $n$.

Simpler questions such as determining just how slow the speed of
convergence of the largest complex
point is to its Gumbel limit would also be interesting.
Working through the proof of Section~\ref{sec3} only produces an $O( (\log
n)^{-1})$ speed estimate. There is no reason
to expect this is close to optimal. On the other hand,
Figure~\ref{fig6}
compares the spectral radius in the $n =100$
complex Ginibre ensemble (in which there are no real points with
probability one) to its corresponding Gumbel limit. The fit is far more
satisfying. Studying the proof from
\citet{MR1986426} of this limit theorem gives an $O (\frac{(\log n)^2}{
\sqrt{n}})$ speed.



%
%

\section*{Acknowledgements}
Many thanks to Torsten Ehrhardt and Harold
Widom for useful discussions.


%

\printaddresses

\end{document}